\documentclass[12pt,a4paper,final]{iopart}

\usepackage{graphicx}
\pdfoutput=1
\usepackage{amsfonts,amsmath,amsthm,amssymb,amsopn}
\usepackage{dsfont,bbm}
\usepackage{mathrsfs}
\usepackage{leftidx}
\usepackage{physics}
\usepackage{cite}
\usepackage{booktabs}
\usepackage[retainorgcmds]{IEEEtrantools}
\usepackage{soul,xcolor}
\usepackage{color}
\definecolor{darkblue}{rgb}{0.,0.,0.5}
\definecolor{darkred}{rgb}{0.6,0.,0.}
\definecolor{darkgreen}{rgb}{0.,0.8,0.}
\definecolor{darkpink}{rgb}{1,0.08,0.5}

\usepackage[breaklinks=true,colorlinks=true,linkcolor=darkblue,urlcolor=magenta,citecolor=magenta]{hyperref}
\numberwithin{equation}{section}
\usepackage{cleveref}
\def\be{\begin{equation}}
\def\ee{\end{equation}}

\newcommand{\ii}{ {\rm i} }

\eqnobysec

\begin{document}
\setstcolor{red}

\title{Fluctuations of stochastic charged cellular automata}
\author{\v{Z}iga Krajnik$^1$, Katja Klobas$^2$, Bruno Bertini$^2$, Toma\v{z} Prosen$^3$}

\address{$^1$Department of Physics, New York University, 726 Broadway, New York, NY 10003, USA}
\address{$^2$School of Physics and Astronomy, University of Birmingham, Edgbaston, Birmingham, B15 2TT, UK}
\address{$^3$Faculty of Mathematics and Physics, University of Ljubljana, Jadranska 19, SI-1000 Ljubljana, Slovenia}

\date{\today}

\begin{abstract}
We obtain the exact full counting statistics of a cellular automaton with freely propagating vacancies and charged particles that are stochastically scattered or transmitted upon collision  by identifying the problem as a colored stochastic six-vertex model with one inert color. Typical charge current fluctuations at vanishing net charge follow a one-parameter distribution that interpolates between the distribution of the charged single-file class in the limit of pure reflection and a Gaussian distribution in the limit of pure transmission. 
\end{abstract}


\maketitle

\pagestyle{empty}
\tableofcontents
\pagestyle{headings}

\flushbottom
\clearpage

\section{Introduction}

Although the study of non-equilibrium many-body dynamics is generally distinct from that of critical phenomena --- it typically involves systems with non-trivial scales for example --- it can nevertheless manifest an astonishing degree of independence from the microscopic details that can be described as `universal'. Remarkable instances of such behavior have been found by looking at \emph{transport phenomena} in \emph{integrable systems}. A famous example is the occurrence of super-diffusive transport~\cite{Znidaric2014a}, with a dynamical structure factor that asymptotically approaches the KPZ scaling function~\cite{Praehofer2004}, in integrable systems with isotropic interactions. The super-diffusive behavior has been theoretically predicted in many different cases and the KPZ scaling inferred numerically 
--- in both the quantum~\cite{Ljubotina2017,Ilievski2018,DeNardis2019,Gopalakrishnan2019,Ljubotina2019a,Dupont2020a,Weiner2020,Denardis2020a,Ilievski2021} and classical~\cite{Das2019,Krajnik2020,Krajnik2020a,McRoberts_Bilitewski_Haque_Moessner_2022,Roy_Dhar_Spohn_Kulkarni_2024} realms --- and also confirmed by experimental observations~\cite{Scheie2021,Wei2022,Jepsen2020}. The study of more refined current fluctuations \cite{Znidaric2014,Krajnik2022a,Krajnik2024a}, however, has revealed that while \emph{universal} and \emph{anomalous} (i.e.~non-Gaussian), this phenomenon is only partially described by the KPZ class~\cite{takeuchi2024, Rosenberg2024}, and the question of what class describes it fully is currently still open. 

More generally, integrable systems with $U(1)$ symmetries have been found to display anomalous charge fluctuations of the same form across a spectrum of systems going from classical~\cite{Krajnik2022,Kormos2022,Krajnik2024,Krajnik2024b} to quantum integrable models~\cite{Krajnik2024a}. This phenomenon was first observed in the microscopic solution~\cite{Krajnik2022} of a simple, reversible cellular automaton describing charged hard-core particles hopping on a one-dimensional lattice~\cite{Medenjak2017} and has been recently embedded in a hydrodynamic framework~\cite{Gopalakrishnan_McCulloch_Vasseur_2024, Yoshimura2024}. In particular --- employing ideas from ballistic macroscopic fluctuation theory \cite{Doyon2023} --- Ref.~\cite{Yoshimura2024} suggested that a similar pattern of universal anomalous fluctuations also occurs in a class of stochastic cellular automata. In this paper, we provide an exact microscopic derivation of this statement.

Specifically, we consider a simple stochastic cellular automaton consisting of freely moving vacancies, and positively and negatively charged hard-core particles that are swapped stochastically upon colliding~\cite{Klobas2018}. Cellular automata and their continuous-time limits often serve as minimal examples to exactly understand diverse classes of many-body dynamical phenomena~\cite{Georges_Le_Doussal_1989,Lebowitz_Maes_Speer_1990}. Indeed, as explicitly shown in the seminal paper~\cite{Gwa_Spohn_1992}, 
one can often use powerful integrability techniques to study their dynamics by mapping them to quantum integrable models. For instance, the exact propagator of the totally asymmetric simple exclusion process in continuous-time for an arbitrary number of particles was obtained in determinant form \cite{Schutz_1997} using Bethe ansatz. 
Moreover, in a series of papers \cite{Tracy_Widom_2008a, Tracy_Widom_2008b, Tracy2009a} analogous results were obtained for the asymmetric exclusion process in terms of multiple contour integrals. Remarkably, while the derivation was purely combinatorial, the result was manifestly of Bethe ansatz form. Recasting the multiple integrals as a Fredholm determinant, the aymtptotics distribution of the right-most particle starting from an step initial condition was obtained, demonstrating that the asymmetric exlcusion process belongs to the Kardar-Parisi-Zhang (KPZ) universality class \cite{Kardar1986,Corwin2012,Takeuchi2018}.
These results were used  \cite{Derrida_Gerschenfeld_2009} to obtain the full counting statistics of the simple symmetric exclusion process (SSEP), microscopically confirming \cite{Derrida2009} a prediction of macroscopic fluctuation theory \cite{Bertini2015a}. The Bethe ansatz structure has been extended to other interacting particle systems \cite{Borodin_Corwin_Petrov_Sasamoto_2015} and to discrete-time \cite{BorodinCorwinGorin2016}. Here we show that the techniques of integrable combinatorics can also be used to study dynamical charge fluctuations.

\subsection{Summary of results}
\label{sec:summary}

\begin{figure}[h!]
\centering
\includegraphics[width=\linewidth]{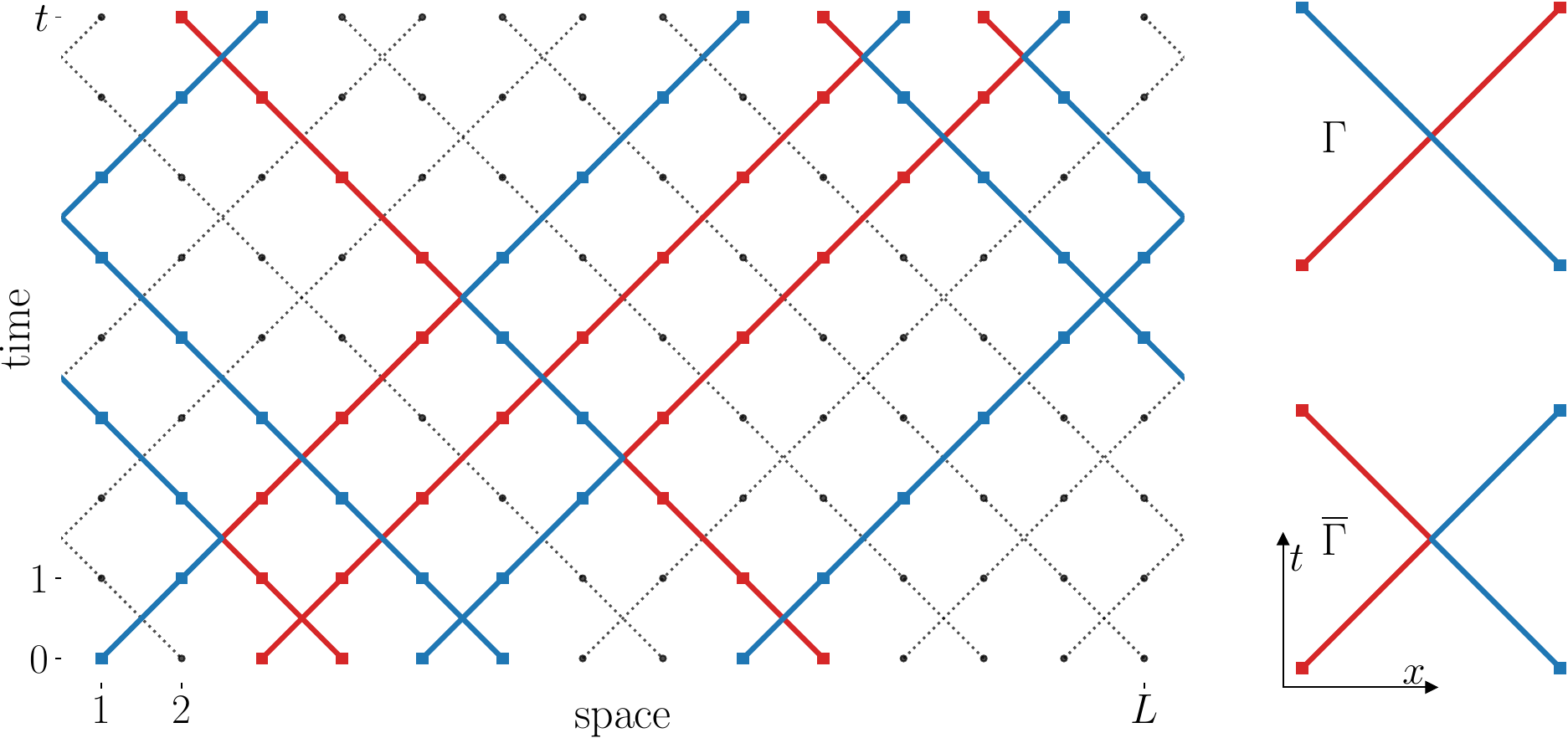}
\caption{Freely propagating vacancies (dashed black lines) and ballistically propagating charged particles  (red/blue lines) on a lattice of length $L$ with periodic boundary conditions. Upon colliding particles are either transmitted with probability $\Gamma$ (upper right panel) or elastically reflected with probability $\overline \Gamma$ (lower right panel).}
\label{fig:dynamics}
\end{figure}

We study an interacting particle model in discrete space-time introduced in Ref.~\cite{Klobas2018}, whose dynamics consist of freely propagating vacancies and charged particles subject to an exclusion rule that allows at most one particle on a given site. Particles propagate ballistically except upon collision, when they are transmitted with probability $0 \leq \Gamma \leq 1$ and elastically reflected with probability $\overline \Gamma \equiv 1-\Gamma$, see Figure \ref{fig:dynamics} for a snapshot of the dynamics.

\noindent We are interested in the probability distribution of the time-integrated charge current
\be
J(t) = \sum_{t'=0}^{\lceil t/2 \rceil} j^c_{1}(2t'),
\ee
where $j^c$ is the charge current density (see Eq.~\eqref{current_def}) defined by a local continuity relation \eqref{local_continuity}. The distribution of $J(t)$ has been found to be anomalous in the limit $\Gamma = 0$~\cite{Krajnik2022,Krajnik2024} and shown to be described by an Euler equation with a fluctuating velocity with zero mean \cite{Yoshimura2024} on the hydrodynamic scale. For simplicity we restrict ourselves to fluctuations in equilibrium.\\

\noindent In Section \ref{sec:model} we define the model's dynamics in terms of sequential product of two-body maps alternatively acting on even and odd pairs of neighboring lattices sites, producing the `brickwork' structure observed in Figure \ref{fig:dynamics}. We define a family of bipartite initial probability measures with fixed particle densities $\rho_\pm$ and charge densities per particle $b_\pm$ to the right (+) and left (-) of the origin. We introduce the exponential generating function of the time-integrated charge current $\langle e^{\lambda J(t)} \rangle$, commonly referred to as the full counting statistics, with respect to this family of measures as the Laplace transform of the corresponding probability distribution $P(J|t)$
\begin{equation}
\langle e^{\lambda J(t)} \rangle = \int_{-\infty}^{\infty} \dd J\,  P(J|t) e^{\lambda J}. \label{FCS_def_intro}
\end{equation}
Interpreting the dynamics of the initial probability measure as a stochastic vertex model and noting that vacancies propagate freely we can contract all vacancy weights of the model and express the full counting statistics as a `vacancy-dressed' full counting statistics of the stochastic six-vertex model
\begin{equation}
\langle e^{\lambda J(t)} \rangle = 
\sum_{n_-, n_+ =0}^t P(n_-|t) P(n_+|t) \mu_-^{n_+ - n_-} \mathbb{E}_{\rm step}(\mu^{N_{t-n_+}}|t-n_-),
 \label{dressed_6_vertex_partition_step_intro}
\end{equation}
where $P(n_\pm|t) =\binom{t}{n_\pm} \rho_\pm^{t-n_\pm} \overline \rho_\pm^{n_\pm}$ with $\bar{\rho}_\pm = 1-\rho_{\pm}$, are the probabilities of having $n_\pm$ vacancies on the left/right boundary of the vertex model and $\mu_\pm = \cosh \lambda \mp b_\pm \sinh \lambda$ are the dressed counting fields with $\mu= \mu_-\mu_+$, while $\mathbb{E}_{\rm step}$ denotes the average with respect to a (bi)stochastic six-vertex model with step initial conditions and $N_x$ counts the number of particles the the left of $x+1$.\\

\noindent In Section \ref{sec:6vert_FCS} we derive a multiple integral representation of the full counting statistics of the bistochastic six-vertex model with step initial conditions, following Borodin, Corwin and Gorin's analysis of the the stochastic six-vertex model \cite{BorodinCorwinGorin2016}
\begin{equation}
\mathbb{E}_{\rm step}(\mu^{N_x}|t)  =  \sum_{k=0}^\infty\frac{(\mu-1)^k}{k!}  \oint_{{c}_r}^{\times k} \prod_{i=1}^k \frac{\dd z_i}{2\pi \ii}   \det \left( \frac{z_i^{t-x}h^t(z_i) }{1 - 2z_i + z_i z_j } \right)_{1\leq i,j \leq k}\label{Estep_weta_final2_intro},
\end{equation}
where ${c}_r$ is a small positively-oriented circle centered at the origin and $h(z) = (1 + (z^{-1} - 2)\Gamma)/({1 - \Gamma z})$ is an exponentiated  `discrete dispersion relation'. An analogous result for the simple symmetric exclusion process has been obtained in Ref.~\cite{Imamura_Mallick_Sasamoto_2021}.
The expression \eqref{Estep_weta_final2_intro} has the form of a Fredholm determinant which enables us to expand the logarithm of the full counting statistics as a series
\begin{equation}
\log \mathbb{E}_{\rm step}(\mu^{N_x}|t) = - \sum^{\infty}_{k=1} \frac{(1-\mu)^k}{k} I_{x, t}^{(k)},
\label{CGF_int_intro}
\end{equation}
in terms of the traces of an integral kernel 
\begin{equation}
I_{x, t}^{(k)}  = \oint^{\times k}_{{c}_r} \prod_{i=1}^k \frac{\dd z_i}{2\pi \ii}  \frac{z_i^{t-x}h^t(z_i)}{1 - 2z_i + z_iz_{i+1}},
\label{trace_def_intro}
\end{equation}
where we identify integration variables in the $k$-th term periodically. To bring the traces \eqref{trace_def_intro} to a form amenable to asymptotic analysis we follow a procedure similar to that employed by Derrida and Gerschenfeld in their study of the simple symmetric exclusion process \cite{Derrida_Gerschenfeld_2009} to obtain
\begin{equation}
 I_{x, t}^{(k)}  =  \Gamma^k   \oint_{{c}_r}^{\times k} \prod_{i=1}^k \sum_{t_i=0}^{t-1}  \frac{\dd z_i}{2\pi \ii z_i}  
  z_i^{t-x} \frac{\left(1 + (z_i^{-1} - 2)\Gamma\right)^{t_i}}{\left(1-\Gamma z_{i}\right)^{t_{i-1}+1}}, \quad {\rm for}\ {t \geq x},
 \label{I_representatio_intro}
\end{equation}
where we identify the summation variables periodically, $t_i = t_{i+k}$.
Since Eq.~\eqref{I_representatio_intro} is only valid for $t \geq x$ we also derive a reflection relation for the full counting statistics
\begin{equation}
\mathbb{E}_{\rm step}(\mu^{N_x}|t)  \mu^t  =  \mathbb{E}_{\rm step}(\mu^{N_t}|x)  \mu^x, \label{step_reflectio_intro}
\end{equation}
that allows us to extend the results obtained using Eq.~\eqref{I_representatio_intro} to all $x$ and $t$.\\

\noindent We start Section \ref{sec:asymptotics}  by evaluating the multiple integrals in the expression \eqref{I_representatio_intro} followed by a resumation to bring it to the form
\begin{equation}
I_{x, t}^{(k)} =
\Gamma^k \sum_{s_1, \ldots, s_k=0}^{x-1} \psi^{(k)}_{t- x}(s_1, \ldots, s_k; \gamma ) \quad {\rm for}\ {t \geq x},\label{pseudo_Vandermond_representation_intro}
\end{equation}
where $\gamma = \overline\Gamma/\Gamma$ and we have introduced the functions $\psi_n^{(k)}$
\begin{equation}
\psi_n^{(k)}(s_1, s_2, \ldots, s_k; \gamma) =  \prod_{i=1}^k  (1+\gamma)^{ - 2s_i-n}  \sum_{j_i=0}^{s_i} \binom{s_i +n}{j_i} \binom{s_{i-1}}{s_i -j_i}   \gamma^{2j_i}.
 \label{psi_def_intro}
\end{equation}
Using Laplace's method we first derive the asymptotic form of $\psi_n^{(k)}$ and, applying it again, we obtain the asymptotic form of the trace sum \eqref{pseudo_Vandermond_representation_intro}
\begin{equation}
I^{(k)}_{x, t} \simeq \sqrt{\frac{t }{\pi \gamma k}} \int_0^1 \dd \xi\, e^{-\frac{k \gamma \delta^2}{4\xi^2} }, \qquad t-x = \delta t^{1/2} \geq 0. \label{trace_localization6_intro}
\end{equation}
Returning to the expression \eqref{CGF_int_intro} and resumming the trace series  we obtain the full counting statistics of a large square bistochastic six-vertex lattice with diffusive shape fluctuations
\begin{equation}
\log \mathbb{E}_{\rm step}(\mu^{N_x}|t) \simeq \sqrt{t\gamma} \int_{0}^1 \dd \xi\, \int_{-\infty}^{\infty}\frac{\dd k}{\pi}\, \log \left(1+(\mu-1) e^{-k^2 - \gamma \delta^2/4 \xi^2} \right) \label{CGF_series_result_intro}
\end{equation}
for $t-x = \delta t^{1/2} \geq 0$. 
We also show that ballistic shape fluctuations are rapidly suppressed, $ \log \left[ \log \mathbb{E}_{\rm step}(\mu^{N_x}|t) \right] \simeq t \Phi^b(\delta) < 0$ for $t-x = \delta t \geq 0$.
\\

\begin{figure}
\centering
\includegraphics[width=\linewidth]{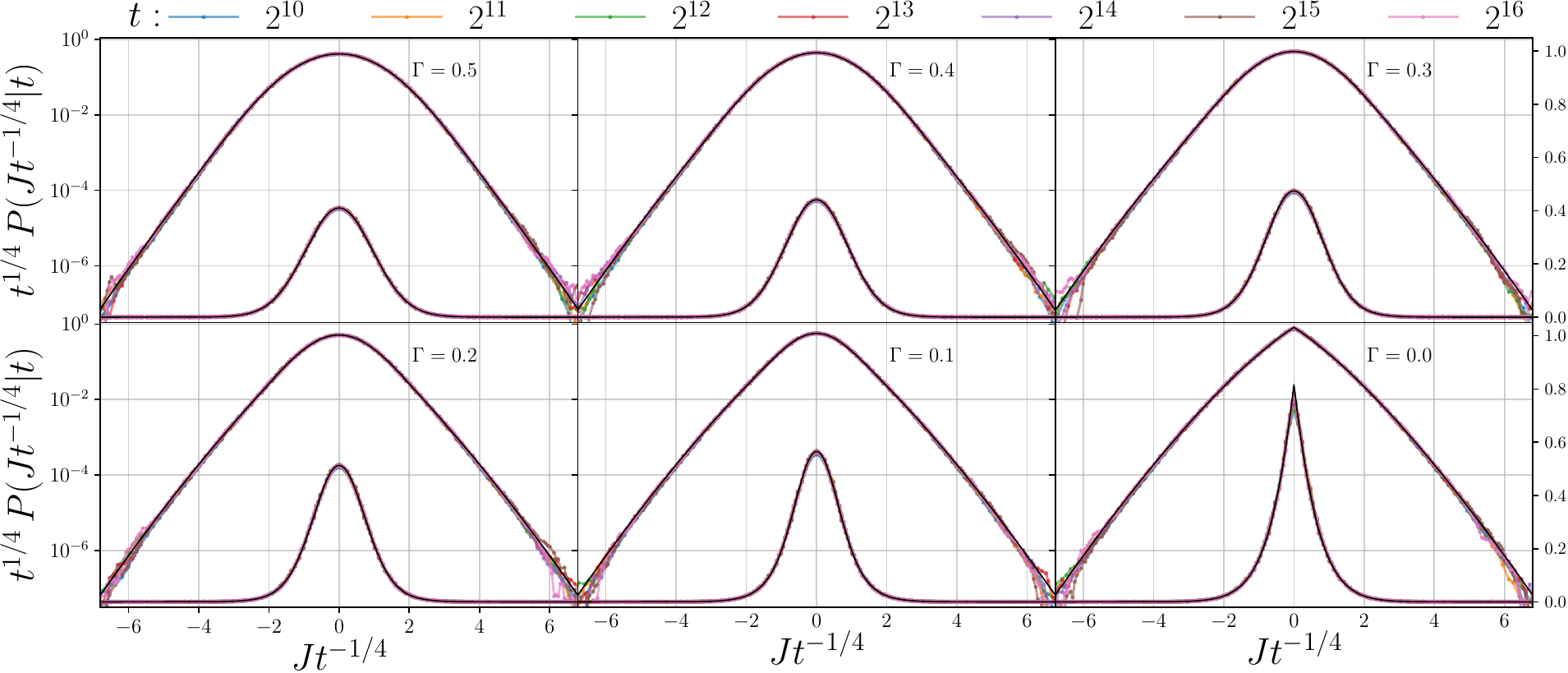}
\caption{Rescaled finite-time distribution of the integrated charge current (colored curves, data shown for $t \geq 2^{10}$) for different values of the crossing parameter $\Gamma$ at zero net charge ($b=0$) in logarithmic and linear scale (top and bottom curves in each panel respectively), compared against the analytical prediction \eqref{typical_final_hf_intro} (black curve). Simulation parameters: $\rho=1/2$, $L= 2^{20}$, averaged over $5\times 10^3$ initial conditions.}
\label{fig:numerical_comparison}
\end{figure}

\noindent In Section \ref{sec:dressed_asymptotics} we use the asymptotic form of the stochastic six-vertex model \eqref{CGF_series_result_intro} to study the vacancy-dressed full counting statistics \eqref{dressed_6_vertex_partition_step_intro} of the charge cellular automaton on both typical and large scales. In equilibrium states with a vanishing charge density typical charge fluctuations occur on the diffusive scale $z=2$ with a one-parameter family of non-Gaussian distributions
\begin{equation}
P_{\rm typ}(j) = \frac{1}{\pi \sigma }\int_{0}^{\infty} \frac{\dd u}{\sqrt{u[1+s(u/a)]}} e^{-\frac{u^2}{2\sigma^2}-\frac{j^2}{2u[1+s(u/a)]}}, 
\label{typical_final_hf_intro}
\end{equation}
where $\sigma^2 = 2  \rho \overline \rho $ and $a = 2\sqrt{\rho /\gamma} \geq 0$ while $s(z) =  \pi^{-1/2} z^{-1}e^{-z^2} - {\rm erfc}\, z$. We compare the analytical prediction \eqref{typical_final_hf_intro} against direct numerical simulation, finding excellent agreement as shown in Figure \ref{fig:numerical_comparison}: this comparison involves no free parameters and is displayed for more than five orders of magnitude. Large charge fluctuations are independent of the crossing probability $\Gamma$ and have a large deviation form
\begin{equation}
 \lim_{t\to \infty} t^{-1}\log \langle e^{\lambda J(t)} \rangle =   \log\left[1 + \rho \overline \rho(\mu_b + \mu^{-1}_b -2) \right],
\end{equation}
where $\mu_b= \cosh \lambda + |b| \sinh|\lambda|$, confirming the recent prediction~\cite{Yoshimura2024} of ballistic macroscopic fluctuation theory.\\

\noindent We conclude in Section \ref{sec:conclusion} by discussing the interpretation of our results  and pointing out some related open questions.

\section{Stochastic  charged cellular automata}
\label{sec:model}

We start by introducing the dynamics of the stochastic cellular automaton of charged particles and vacancies of interest in this work, see also Ref.~\cite{Klobas2018}. The model can be understood as a stochastic six-vertex model with three colors, with one of the colors --- termed vacancy --- having trivial free dynamics. We define a family of bipartite initial probability measures with fixed average charge and particle densities, and the full counting statistics of the charge current. Using the simple vacancy dynamics we express the full counting statistics as those of a vacancy-dressed stochastic six-vertex model.

\subsection{Dynamics}
The configuration space $\mathcal{C}_L$ of a discrete system of $L \in \mathbb{N}$ sites is spanned by strings
\begin{equation}
	{\bf s} = s_{1}s_{2} \ldots s_L
\end{equation}
of symbols $s_x$ representing the state of a given site. In our case we consider $s_x \in \{\emptyset, +, -\}$, which correspond to empty sites (vacancies) and negative/positive particles respectively. The local dynamics of the symbols are given by a one-parameter stochastic two-body map $\Phi: \mathcal{C}_2 \to \mathcal{C}_2$ which encodes the following dynamical rules involving either at least one vacancy
\begin{equation}
	(\emptyset, \emptyset) \rightarrow (\emptyset, \emptyset), \qquad (\emptyset, s) \leftrightarrow (s, \emptyset),\\
\end{equation}
or two charged particles
\begin{equation}
	 (s, s') \rightarrow
	\begin{cases}
	(s', s), &\textrm{with probability $\Gamma$},\\
	(s, s'), &\textrm{with probability $\overline \Gamma$},
	\end{cases} \qquad s, s' \in \{-, +\}. \label{prop_def}
\end{equation}
Here $ 0 \leq \Gamma \leq 1$ and $\overline \Gamma \equiv 1-\Gamma$ are respectively the crossing and reflection probabilities. The local  particle dynamics is embedded into the many-body configuration space $\mathcal{C}_L$ as
\begin{equation}
	\Phi_{x, x+1} = {\rm Id}^{\otimes (x-1)} \otimes \Phi \otimes {\rm Id}^{\otimes (L-x-1)}, \label{embedding}
\end{equation}
where ${\rm Id}(s) = s$ is a local unit map and we impose periodic boundary conditions by identifying site indices as $x \equiv x+L$. The full propagator implementing the many-body dynamics
of a system of even length $L \in 2\mathbb{N}$ consists of alternating even and odd steps
\begin{equation}
	\Phi^{\rm full} = \Phi^{\rm odd} \circ \Phi^{\rm even}, \label{full_def}
\end{equation}
where $\circ$ denotes a composition of maps. The even and odd steps further decompose as
\begin{equation}
	\Phi^{\rm odd} = \prod_{x=1}^{L/2} \Phi_{2x-1, 2x}, \qquad \Phi^{\rm even} = \prod_{x=1}^{L/2} \Phi_{2x, 2x+1}.
\end{equation}
The $t$-step propagator is then given by
\begin{equation}
\Phi^t = 
\begin{cases}
[\Phi^{\rm full}]^{\frac{t}{2}} & {\rm for}\ t \in 2\mathbb{N},\\
\Phi^{\rm even} \circ [\Phi^{\rm full}]^{\frac{t-1}{2}} & {\rm for}\ t \in 2\mathbb{N}-1,
\end{cases} \label{t_step_propagator}
\end{equation}
The dynamics \eqref{t_step_propagator} for $\Gamma=0$ and $\Gamma=1$ are deterministic: in the former case particles have a hard core constraint and a given initial ordering is maintained at all times, while in the latter the particles are fully non-interacting. We refer to these cases respectively as `single-file' and `free' dynamics. Instead of propagating strings ${\bf s} \in \mathcal{C}_L$, we can alternatively propagate observables $O: \mathcal{C}_L \to \mathbb{R}$ according to
\begin{equation}
{O}(t) = {O} \circ \Phi^t.
\end{equation}
We order the local basis $| s \rangle$ of symbols as $(\emptyset, +, -)$. An arbitrary many-body measure $\varrho$ can be expanded along the product basis as
\begin{equation}
| \varrho \rangle = \sum_{{\bf s} \in \mathcal{C}_L} \varrho_{\bf s}  | {\bf s}\rangle,
\end{equation}
where $\varrho_{\bf s} = \langle {\bf s} | \varrho \rangle$ are coefficients of the measure along the basis vector $| {\bf s} \rangle = \prod_{x=1}^L | s_x \rangle$.
 We henceforth work in the limit of large systems by sending $L \to \infty$. \\
 
\begin{figure}[h!]
    \centering
    \includegraphics[width=\linewidth]{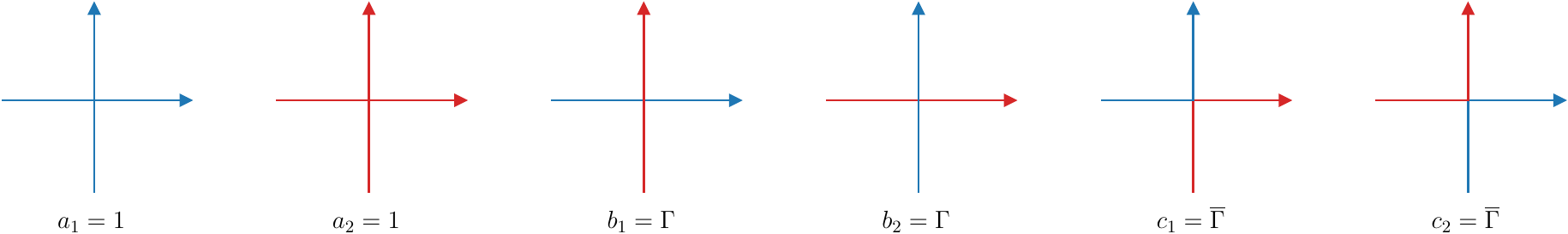}
    \caption{The particle-particle (blue/red lines) sector of the two-body map $\Phi$ \eqref{six_vertex} corresponds to a bistochastic six-vertex model upon identifying vertex weights as shown above.}
    \label{fig:weights}
\end{figure}

\noindent As an aside, we note that ordering the two-body basis of symbols $ | s_x s_{x+1} \rangle$ as $(\emptyset\emptyset, \emptyset+, \emptyset-, +\emptyset, -\emptyset, ++, +-, -+, --)$ the local two-body map takes the form of a $3^2 \times 3^2$  bistochastic matrix\footnote{We henceforth abuse notation by using the same symbol to represent a map on the configuration space and its matrix representation that acts on the basis vectors $|{\bf s} \rangle$.}
\begin{equation}
	\langle s^{t+1}_{x} s^{t+1}_{x+1} |\Phi | s^t_{x} s^t_{x+1} \rangle = 
	\begin{bmatrix}
	1 & & & \\
	& & \mathds{I} &	\\
	& \mathds{I} & & \\
	& & & U 
	\end{bmatrix}, \label{prop_mat}
\end{equation}
where $\mathds{I} = {\rm diag}(1, 1)$ and $U$ is the matrix of the  (bi)stochastic six-vertex model 
\begin{equation}
\langle s_x^{t+1} s_{x+1}^{t+1} |U |s_x^t s_{x+1}^t \rangle = 
\begin{bmatrix}
	1 & &  & \\
	& \overline\Gamma  &  \Gamma  & \\
	&  \Gamma & \overline \Gamma & \\	
	& & & 1 
\end{bmatrix}, \label{six_vertex}
\end{equation}
where $s \in \{+, - \}$.
The two-body map can therefore be understood as the combination of a stochastic six-vertex model in the particle-particle subspace (given by Eq.~\eqref{six_vertex}) and a swap (i.e.~free dynamics) in the vacancy-vacany and vacany-particles subspaces (i.e.~the structure observed in Eq.~\eqref{prop_mat}). In terms of the standard six-vertex weight notations the matrix $U$ corresponds to $a_1 = a_2 = 1$, $b_1 = b_2 =\Gamma$ and $c_1=c_2=\overline \Gamma$, see Figure \ref{fig:weights}. We note that the weights satisfy the stochasticity condition $(a_1-b_1)(a_2-b_2) = c_1 c_2$.

\subsection{Initial measures}
We consider initial probability measures that are \emph{bipartite}, \emph{uniform} on the two halves of the system, and \emph{factorised} in terms of one-site measures. Namely~\footnote{We use the convention that  ${\rm sgn}\ 0 = -1$.}
\begin{equation}
	| \varrho_{\rm ini} \rangle =  \bigotimes_{x = -\infty}^{\infty}|\varrho_{{\rm sgn}\, x} \rangle,
	 \label{measure_def}
\end{equation}
with  
\begin{equation}
|\varrho_\pm \rangle = \left(\overline \rho_\pm,  \rho_\pm \tfrac{1+ b_\pm}{2}, \rho_\pm \tfrac{1- b_\pm}{2} \right)^T. 
\label{rho_pm}
\end{equation}
Here $0 \leq \rho_\pm \leq 1$, $\overline \rho_\pm \equiv 1 - \rho_\pm$ and $ -1 \leq b_\pm \leq 1$ the densities of particles, vacancies and charge per particle to the right and left of the origin respectively. More explicitly, introducing the local particle number ${p}: \mathcal{C}_1 \to \mathbb{R}$ and charge ${c}: \mathcal{C}_1 \to \mathbb{R}$ observables as
\begin{equation}
{p}(\emptyset) = 0, \qquad
{p}(\pm) = 1, \qquad 
{c}(\emptyset) = 0, \qquad 
{c}(\pm) = \pm 1
\end{equation}
their actions on the one-site measures \eqref{rho_pm} read
\begin{equation}
{p} |\varrho_\pm \rangle = \rho_\pm, \qquad {c} |\varrho_\pm \rangle = \rho_\pm b_\pm. 
\end{equation}
By embedding ultralocal observables $o$ into the many-body space as in Eq.~\eqref{embedding}
\begin{equation}
o_x = {\rm Id}^{\otimes (x-1)} \otimes {o}  \otimes {\rm Id}^{\otimes (L-x)}
\end{equation}
we also define the total charge $C$ and particle number $P$
\begin{equation}
C = \sum_{x=-\infty}^\infty {c}_{x}, \qquad P = \sum_{x=-\infty}^\infty {p}_{x},
\end{equation}
both of which are conserved by the dynamics \eqref{t_step_propagator}
\begin{equation}
C = C \circ \Phi^t, \qquad P = P \circ \Phi^t. 
\end{equation}
Note that these are far from being the only conserved charges of the system. An extensive number of additional linearly independent local conserved quantities can be obtained by considering the number of particles on an arbitrary set of left/right running diagonals. The simplest example is the conservation of all left/right moving particles, see e.g.~Ref.\cite{Yoshimura2024}.\\

\noindent The local densities of $P$ and $C$ satisfy a discrete continuity equation of the form
\begin{equation}
	o_{x, t+2} - o_{x, t} + j^o_{x+1, t+1} - j^o_{x, t+1} = 0, \qquad o \in \{p, c \}\label{local_continuity}, 
\end{equation}
with local current densities
\begin{equation}
	j^o_{x, t+1} = (-1)^x \left(o_{x, t+1+(-1)^x}  -  o_{x, t+1} \right), \qquad o \in \{p, c \}. \label{current_def}
\end{equation}
The average of an observable ${O}:\mathcal{C}_\infty \to \mathbb{R}$ with respect to the measure \eqref{measure_def} reads
\begin{equation}
	\langle {O} \rangle 
	= \langle \varrho_{\rm flat} | O | \varrho_{\rm ini} \rangle, 
\end{equation}
where we have introduced the (unnormalized) flat state as a product of local flat states $| \mathds{1} \rangle = (1, 1, 1)^T$
\begin{equation}
	| \varrho_{\rm flat} \rangle = \bigotimes_{\ell = -\infty}^{\infty}| \mathds{1} \rangle.
\end{equation} 
We note that when the initial measure \eqref{measure_def} is uniform on the whole system i.e.~for
$\rho_\pm = \rho$ and $b_\pm = b$, it is invariant under the many-body dynamics.

\subsection{Full counting statistics and a dressed bistochastic six-vertex model}
\begin{figure}
    \centering
    \includegraphics[width=\linewidth]{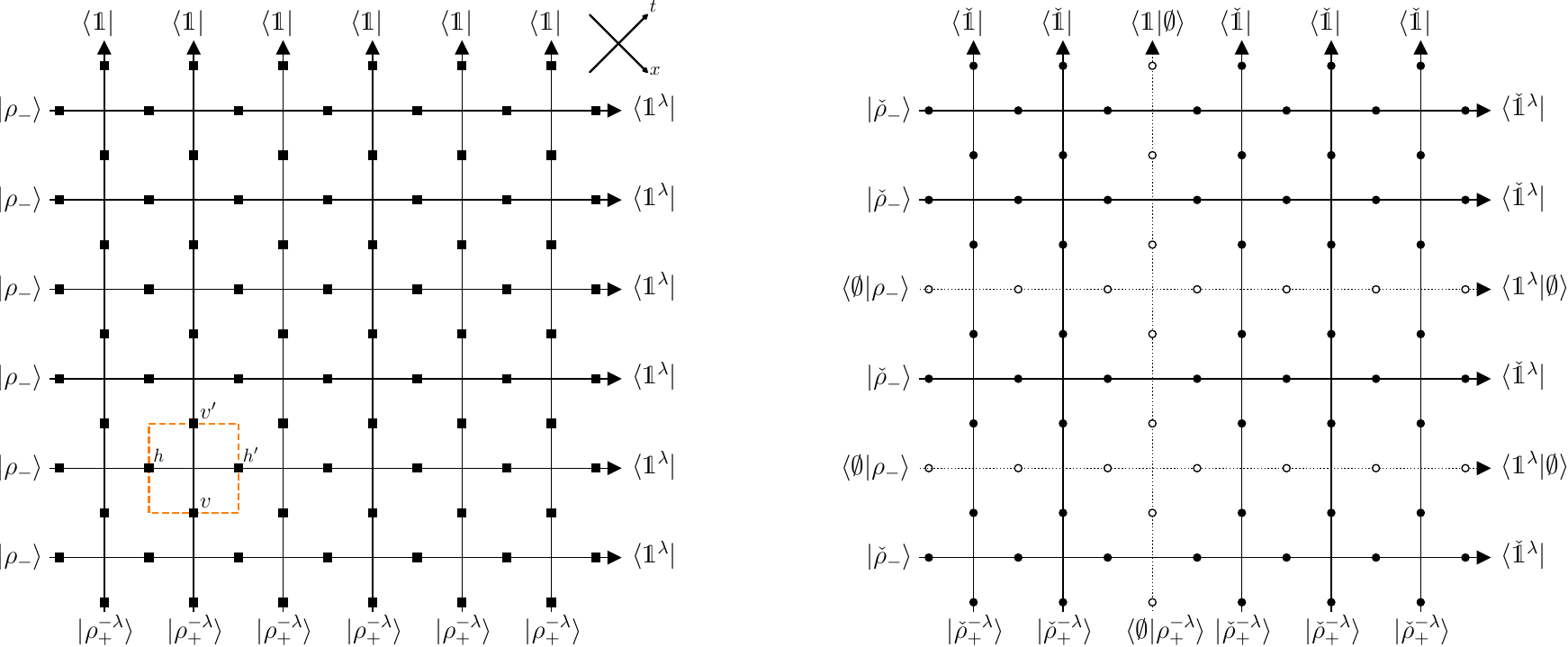}
    \caption{(left panel) The $t\times t$ partition sum of the vertex model \eqref{11_vertex_partition} obtained by contracting the full-counting statistics \eqref{G_lambda_full} (for $t=6$) using the relation \eqref{Phi_conservation}. Dashed orange square indicates the elementary vertex $\Phi_{\,h\, v}^{h' v'}$ defined in Eq.~\eqref{Phi_vertex} with corresponding vertical and horizontal indices (black squares) taking values in $\{\emptyset,+, -\}$. Axes in the top right corner indicate space-time directions of the underlying cellular automaton. (right panel) A contribution to the sum \eqref{6_vertex_partition} obtained after eliminating $n_+ =1$ and $n_-=2$ vacancies (empty circles) from respectively the horizontal and vertical boundaries of the partition sum \eqref{11_vertex_partition} using the relation \eqref{vac_vertex}. The remaining $t-n_+\times t-n_-$ partition sum (black circles take values in $\{+, -\}$ while the elementary vertex is defined in Eq.~\eqref{V_vert_def}) is identified with the stochastic six-vertex model with domain wall-boundary conditions after accounting for the overlaps of the boundary states, see Eq.~\eqref{dressed_6_vertex_partition_step}.}
    \label{fig:vertex_model}
\end{figure}

We are interested in the integrated charge current across the origin
\begin{equation}
J(t) = \sum_{t'=0}^{\lceil t/2 \rceil} j^c_{1}(2t').
\end{equation}
which equals the difference of charges on the right half of the system $C_+ = \sum_{x=1}^\infty c_x$ evaluated at the initial and final times via the continuity equation \eqref{local_continuity}
\begin{equation}
J(t) = C_+(t) - C_+(0). \label{Jc_half}
\end{equation}
The distribution of the integrated charge current $P(J|t)$ is encoded in the exponential generating function, also known as the the full counting statistics 
\begin{equation}
\langle e^{\lambda J(t)} \rangle = \int_{-\infty}^{\infty} \dd J\,  P(J|t) e^{\lambda J}, \label{FCS_def}
\end{equation}
where $\lambda \in \mathbb{C}$ is referred to as the counting field. Using Eq.~\eqref{Jc_half} the full counting statistics becomes
\begin{equation}
	\langle e^{\lambda J(t)} \rangle= \langle \varrho_{\rm flat} |e^{\lambda C_+} \Phi^t e^{-\lambda C_+} | \varrho_{\rm ini}\rangle. \label{G_def}
\end{equation}
The exponentials of the counting fields can be absorbed into the states
\begin{equation}
	\langle e^{\lambda J(t)} \rangle= \langle \varrho^\lambda_{\rm flat} |\Phi^t | \varrho^{-\lambda}_{\rm ini}\rangle. \label{G_lambda_full}
\end{equation}
by introducing a $\lambda$-twist of a many-body measure $|\varrho_{\rm mb} \rangle = \bigotimes_{x = -\infty}^{\infty}|\varrho \rangle$ as
\begin{equation}
	|\varrho^\lambda_{\rm mb} \rangle = \bigotimes_{x = -\infty}^{\infty} e^{\theta(x) \lambda c}|\varrho \rangle,  \label{twist_def}
\end{equation}
with the matrix form of the charge $c = {\rm diag}(0, 1, -1)$ and the Heaviside function
\begin{equation}
\theta(x) =
 \begin{cases}
0 & x \leq 0,\\
1 & x > 0.
\end{cases}
\end{equation}
We now observe that the two-body map $\Phi$ acts as an identity on a tensor product of two arbitrary identical one-site measures
\begin{equation}
|\varrho \rangle \otimes |\varrho \rangle = \Phi |\varrho \rangle \otimes |\varrho \rangle. \label{Phi_conservation}
\end{equation}
Repeated applications of Eq.~\eqref{Phi_conservation} allows us to contract the expression \eqref{G_lambda_full} to a $t \times t$ partition sum of the vertex model defined by the map $\Phi$, see the left panel of Fig.~\ref{fig:vertex_model}
\begin{equation}
\langle e^{\lambda J(t)} \rangle =   \langle \mathds{1} |{}^{\otimes t} \langle\mathds{1}^\lambda | M | \varrho_-\rangle^{t} | \varrho_+^{-\lambda} \rangle^{\otimes t},\label{11_vertex_partition}
\end{equation}
where we have introduced the monodromy matrix $M$ 
\begin{equation}
 M_{\, h_1 v_1\ldots v_t}^{h_{t+1} v_1' \ldots v_t' }  = \Phi_{h_t, v_t}^{h_{t+1},v_t'} \ldots \Phi_{h_1 v_1}^{h_2 v_1'}. \label{M_def1}
\end{equation}

\noindent In Eq.~\eqref{M_def1} and hereafter we use the summation convention of summing over repeated indices.
The vertex $\Phi_{h\, v}^{h' v'}$ encodes the two-body map $\Phi$ according to the identification\footnote{Note that $\Phi_{h\, v}^{h' v'}$ is defined in Eq.~\eqref{Phi_vertex} and should not be confused with the standard notation for a matrix element, i.e.~$\Phi_{\,h\, v}^{h' v'} \neq \langle h' v' | \Phi | h v \rangle$.}
\begin{equation}
\Phi_{h\,\, v}^{h' v'}  = \langle v' h' | \Phi | h v \rangle, \label{Phi_vertex}
\end{equation}
for $h, v \in \{\emptyset, +, - \}$. A comment on the notation used in Eq.~\eqref{11_vertex_partition} is in order - the top/bottom boundary states of the partition function (see Figure~\ref{fig:vertex_model})  consist of $t$ tensor products of their respective one-site states. On the other hand, each horizontal line consists of the monodromy matrix $M$ \eqref{M_def1} contracted on the left/right with a one-site state. Since the partition function consists of $t$ horizontal lines, the contracted monodromy matrix is raised to the $t$-th power which already accounts for the left/right boundary states. We note that this is consistent with the identification in Eq.~\eqref{Phi_vertex}.\\

Since vacancies move freely, the vertex acts trivially on both a horizontal or a vertical vacancy
\begin{equation}
\Phi_{\emptyset\, \,  v}^{h' v'} = \delta_{h', \emptyset}, \qquad  \Phi_{h\,\,v}^{h' \emptyset}  = \delta_{\emptyset, v} \label{vac_vertex}.
\end{equation}
These  relations allow us to easily contract all the vacancy-subspaces in the partition sum \eqref{11_vertex_partition}. To see this we insert a resolution of the identity 
$\mathds{1} = |\emptyset\rangle \langle \emptyset| + |+\rangle \langle+| + |-\rangle \langle - |$ at the bottom horizontal and left vertical boundaries of the vertex model. We then expand the partition sum into $4^t$ contributions by choosing either a vacancy or a positive/negative particle at each resolution. Since vacancies propagate freely by Eq.~\eqref{vac_vertex} the entire corresponding row or column can be contracted, see right panel of Fig.~\ref{fig:vertex_model}.
One then observes that, crucially, the contributions do not explicitly depend on the position of the vacancies, but only on their number on the vertical and horizontal boundaries. Specifically, denoting the latter by $n_+$ and $n_-$ we find
\begin{equation}
\hspace{-10pt}\langle e^{\lambda J(t)} \rangle = \hspace{-10pt} \sum_{n_-,n_+ =0}^t \binom{t}{n_-}\binom{t}{n_+}  \langle \emptyset | \varrho_+^{-\lambda} \rangle^{n_+} \langle \emptyset | \varrho_- \rangle^{n_-}  \langle \check {\mathds{1}} |^{\otimes t - n_-} \langle \check {\mathds{1}}^{\lambda} | \mathcal{M}| \check \varrho_-\rangle^{t-n_+}   | \check \varrho_+^{-\lambda} \rangle^{\otimes t-n_-} \label{6_vertex_partition}
\end{equation}
where $\mathcal{M}$ is the monodromy matrix of the stochastic six-vertex model
\begin{equation}
 \mathcal{M}_{\,h_1 v_1\ldots v_t}^{h_{t+1}v_1' \ldots v_t'}  =  U_{\,\, h_t, v_t}^{h_{t+1},v_t'} \ldots U_{\,\, h_1 v_1}^{h_2 v_1'},
\end{equation}
and $\check \bullet$ denotes the corresponding object in the $\{+, -\}$ subspace. Explicitly we have
\begin{equation}
| \check{\mathds{1}} \rangle = (1, 1)^T, \qquad  |\check \varrho_\pm \rangle = \left(\rho_\pm \tfrac{1+ b_\pm}{2}, \rho_\pm \tfrac{1- b_\pm}{2} \right)^T \label{rho_pmrescaled},
\end{equation}
while the $\lambda$-twist \eqref{twist_def} now involves multiplication with $e^{\lambda \check c}$ where $\check c = {\rm diag}(1, -1)$.
The vertex $U_{\, \, h\,\,  v}^{h' v'}$ is obtained from the six-vertex matrix $U$ \eqref{six_vertex} as
\begin{equation}
U_{\, h\,\,  v}^{h' v'}  = \langle v' h' | U  | h v \rangle, \label{V_vert_def}
\end{equation}
where now $h, v \in \{+, -\}$. To further simplify the expression \eqref{6_vertex_partition} we now observe that the six-vertex matrix $U$ maps any tensor  product of two one-site measures as
\begin{equation}
	U | \check \varrho \rangle \otimes | \check \varrho' \rangle = \overline \Gamma | \check \varrho \rangle \otimes | \check \varrho' \rangle + \Gamma |\check \varrho' \rangle \otimes |\check \varrho\rangle, \label{swap_property}
\end{equation}
which allows us to map the evaluation of the six-vertex partition function in \eqref{6_vertex_partition} to the partition function of a six-vertex model with step initial conditions with horizontal and vertical dimensions $t-n_+$ and $t-n_-$ respectively. To demonstrate this we compute the combinatorial weights corresponding to a given number of particle crossing from the bottom to the top boundary in a given interval. We introduce a basis labeled by $v \in \{\uparrow, \downarrow \}$ with the normalization $\langle v | v' \rangle = \delta_{v, v'}$,
identify 
\begin{equation}
| \check \varrho_+^{-\lambda} \rangle \mapsto |\uparrow\ \rangle, \qquad | \check \varrho_- \rangle \mapsto |\downarrow\ \rangle \label{identification}
\end{equation}
and introduce an observable $N_x$ that counts the number of up arrows to the left of the position $x+1$ on a horizontal line of the six-vertex model 
\begin{equation}
N_x = \sum_{x'=1}^x \langle v_{x'} | \uparrow \rangle.
\end{equation}
A direct inspection of the six-vertex partition sum in \eqref{6_vertex_partition} now shows that a configuration with a fixed number $n$ of $|\uparrow\ \rangle$ passing from the bottom to the top boundary produces the following product of boundary overlaps
\begin{equation}
\langle \check{\mathds{1}} | \check \varrho_+^{-\lambda} \rangle^{n} \langle \check{\mathds{1}} | \check \varrho_- \rangle^{t-n_+ -n} \langle \check{\mathds{1}}^\lambda | \check \varrho_+^{-\lambda} \rangle^{t-n_+ -n} \langle \check{\mathds{1}}^\lambda | \check \varrho_- \rangle^{n + n_+ -n_-}.
\end{equation}
We evaluate all the required overlaps while recalling the identification \eqref{identification}
\begin{align}
&\langle \emptyset | \varrho_+^{-\lambda} \rangle = \overline \rho_+, \qquad
\ \, \langle \emptyset | \varrho_- \rangle = \overline \rho_-, \qquad 
\ \, \langle \check{\mathds{1}} | \check \varrho_+^{-\lambda} \rangle  = \rho_+ \mu_+(\lambda),\\
&\ \langle \check{\mathds{1}} | \check \varrho_- \rangle = \rho_-, \quad \enspace \,
\langle \check{\mathds{1}}^\lambda | \check \varrho_+^{-\lambda} \rangle = \rho_+, \qquad \enspace
\langle \check{\mathds{1}}^\lambda | \check \varrho_- \rangle = \rho_- \mu_-(\lambda),
\end{align}
where we have introduced the dressed counting fields
\begin{equation}
\mu_\pm(\lambda) = \cosh \lambda \mp b_\pm \sinh \lambda.
\end{equation}
Further introducing the reduced variable
\begin{equation}
\mu(\lambda) = \mu_+(\lambda) \mu_-(\lambda),
\label{mu_def}
\end{equation}
and collecting all the terms, the full counting statistics \eqref{6_vertex_partition} is equivalent to
\begin{equation}
\langle e^{\lambda J(t)} \rangle = \sum_{n_-,n_+ =0}^t P(n_-|t) P(n_+|t) \mu_-^{n_+ - n_-} \mathbb{E}_{\rm step}(\mu^{N_{t-n_+}}|t-n_-),
 \label{dressed_6_vertex_partition_step}
\end{equation}
where $P(n_\pm|t) =\binom{t}{n_\pm} \rho_\pm^{t-n_\pm} \overline \rho_\pm^{n_\pm}$ are the probabilities of having $n_\pm$ vacancies on their respective boundaries (see right panel of Fig.~\ref{fig:vertex_model})
and $\mathbb{E}_{\rm step}(\mu^{N_{x}}|t)$ is the generalized full counting statistics of the stochastic six-vertex model  that encodes the combinatorial weight for a given number of crossings starting at time $t$ starting from a step (i.e. domain wall) initial conditions
\begin{equation}
\mathbb{E}_{\rm step}(\mu^{N_{x}}|t) = \sum_{n=0}^{\infty} \mu^n \sum_{0 \leq x_1 < \ldots < x_n \leq x} \left[\mathcal{T}^t_x\right]_{ \uparrow\  \ldots\  \uparrow}^{\uparrow_{(x_1 \ldots x_n)}}. \label{E_FCS_def}
\end{equation}
where $\uparrow_{(x_1 \ldots x_n)}$ denotes the string of symbols consisting of $\uparrow$ at positions $x_1, \ldots, x_n$ with the complement being $\downarrow$ symbols and $\mathcal{T}^t_x$ is the $t$-th power of the following transfer matrix of length $x$
\begin{equation}
\left[\mathcal{T}_x\right]_{v_1\ldots v_x}^{v_1' \ldots v_x'} = \sum_{h' \in \{\uparrow, \downarrow\}}\mathcal{M}_{\downarrow\ v_1\ldots v_x}^{h' v_1' \ldots v_x'}, 
\label{T_finite_def}
\end{equation}
with indices $v \in \{\uparrow, \downarrow \}$.\\


\noindent The full counting statistics \eqref{dressed_6_vertex_partition_step} therefore take the form of a `vacancy-dressed' stochastic six-vertex model or alternatively as that of a three-color stochastic six-vertex models with one non-interacting color.
We note that the existence of an analogous reduced variable $\mu$ \eqref{mu_def} has already been demonstrated for SSEP in Ref.~\cite{Derrida_Gerschenfeld_2009}, see also Ref.~\cite{Derrida_Doucot_Roche_2004}.\\

\section{Full counting statistics of the bistochastic six-vertex model}
\label{sec:6vert_FCS}

In this section we obtain an exact expression for the full counting statistics of the bistochastic six-vertex model starting from a step initial condition in the form of a Fredholm determinant. Our derivation and notation closely follow the analysis of the stochastic six-vertex model by Borodin, Corwin and Gorin~\cite{BorodinCorwinGorin2016}. 

\subsection{Preliminaries}
The dynamics of the stochastic six-vertex model is generated by the infinite transfer matrix
\begin{equation}
\left[\mathcal{T}_\infty\right]_{V}^{V'} = \overleftarrow{\prod_{x \in \mathbb{Z}}} U_{\, h_x \enspace \ v_x}^{h_{x+1} v_x'},
\label{T_infinite_def}
\end{equation}
where $V = \ldots v_{-1} v_0 v_1 \ldots$ is an infinite string of symbols $v_x \in \{\uparrow, \downarrow\}$ that describes the incoming (vertical) configuration and, analogously, $V'$ describes the outgoing one. In this section we refer to $\uparrow$ and $\downarrow$ as particles and holes respectively. To each configuration $V$ with $N$ particles we associate the ordered sequence of positions $X=(x_1, x_2, \ldots, x_N) \in \mathbb{W}_N \subset \mathbb{Z}^{\times N}$ of the particles with $x_{i} < x_{i+1}$ for all $1 \leq i \leq N$. We denote the transfer matrix that maps $Y$ to $X$ as $\mathcal{T}_\infty(Y\mapsto X)$.
To extend the definition to an infinite number of particles we consider \emph{stabilizing sequences} $X \in \mathbb{W}_\infty$ for which $x_{i+1} = x_{i} + 1$ for all $i \geq i_{\rm stab}$. Note that, for $\Gamma < 1$, a transition amplitude between two sequences $Y$ and $X$ with an infinite number of particles is non-zero only when $y_{i}=x_{i}+1$ for all large enough $i$.  
The step initial condition is given by the configuration $V = \ldots \downarrow\uparrow \ldots$ with the leftmost particle at $x_1=1$ which corresponds to the stabilizing sequence $X = (1, 2, \ldots)$. The restriction to stabilizing sequences and initial states whose leftmost particle is to the right of the origin (i.e.~$x_1 \geq 1$)
makes the definition \eqref{T_infinite_def} compatible with the finite transfer matrix defined in \eqref{T_finite_def}, see proof of Theorem 4.9 in Ref.~\cite{BorodinCorwinGorin2016}. Intuitively, this follows from the fact that one can obtain the finite transfer matrix \eqref{T_finite_def} from the infinite one \eqref{T_infinite_def} by noting that the (unique) fixed point of the transfer matrix propagating in the $x$ direction is the tensor product of local flat states (sum of $\uparrow$ and $\downarrow$).\\

\noindent The average value of an observable $O: \mathbb{W}_N \to \mathbb{R}$ with respect to an initial sequence $Y\in \mathbb{W}_N$ with $N$ particle at time $t$ is then given by
\begin{equation}
\mathbb{E}_Y(O) = \sum_{X \in \mathbb{W}_N} \mathcal{T}_\infty^t(Y \mapsto X) O(X).
\end{equation}

\subsubsection{Symmetrization identities} In the upcoming derivation we will use a number of symmetrization identities which we list here for convenience.
We define the symmetrization of a $k$-variable function $g$ as the sum over all the permutations of its arguments
\begin{equation}
{\rm sym}[g(z_1, z_2, \ldots, z_k)] = \sum_{\sigma \in \mathfrak{S}_k} g(z_{\sigma(1)}, z_{\sigma(2)}, \ldots, z_{\sigma(k)}), 
\label{sym_def}
\end{equation}
where $\mathfrak{S}_k$ is the symmetric group of $k$ elements.
We note that multiplication with products of single variable functions commutes with the symmetrization operation
\begin{equation}
{\rm sym}\left[g(z_1, z_2, \ldots, z_k) \right] \prod_{i=1}^k f(z_i) = {\rm sym} \left[g(z_1, z_2, \ldots, z_k)  \prod_{i=1}^k f(z_i)\ 
\right].
\label{sym_factor}
\end{equation}
We have the following pair of symmetrization identities
\begin{align}
&{\rm sym} \hspace{-3pt}\left[\prod_{1\leq i < j \leq k} \frac{z_j - z_i} {1-2 z_i + z_i z_j} \prod_{i=1}^k \frac{1- z_i}{ z_i \ldots z_k - 1} \right]
\hspace{-5pt} = \hspace{-3pt}\frac{(-1)^k}{k!}{\rm sym}\hspace{-3pt} \left[ \prod_{1 \leq i < j \leq k} \frac{z_j - z_i} {1-2 z_i + z_i z_j}\right],\label{sym2_equation} \\
&{\rm sym}\left[ \prod_{1\leq i < j \leq k} \frac{1-2 z_i +  z_i z_j}{z_j - z_i}  \right] = k!,\, 
\label{sym_eq_fac1}
\end{align}
see Ref.~\cite{BorodinCorwinGorin2016} and references therein. We also use a determinant relation due to Tracy and Widom \cite{Tracy_Widom_2008a}, see also \cite{Derrida_Gerschenfeld_2009}
\begin{align}
\left[\prod_{i=1}^k \frac{1}{(1-z_i)^2} \right]\left[\prod_{i\neq j} \frac{z_i - z_j}{1 + z_i z_j  - 2z_i} \right] &= \det \left( \frac{1}{1 + z_i z_j - 2z_i} \right)_{1\leq i, j \leq k}.  \label{TW_det_sum}
\end{align}

\subsection{Multiple integral representation}
Let $P_Y(x_n=x|t)$ denote the probability that, starting from an arbitrary initial sequence $Y$ of $N$ particles, the $n$-th particle, with $1\leq n \leq N$, is at position $x$ at time $t$.  Our starting point is a multiple-integral representation of $P_Y(x_n=x|t)$ obtained in \cite{BorodinCorwinGorin2016}, see Theorem 4.9. Specializing to the bistochastic six-vertex model we have
\begin{align}
&P_Y(x_n=x|t) = (-1)^{n-1} \sum_{n\leq k\leq N}\binom{k-1}{n-1} \sum_{|S|=k}\ \nonumber  \\
& \oint_{c_R}^{\times k} \prod_{\stackrel{i,j \in S}{i<j}} \frac{z_j-z_i}{1-2z_i + z_i z_j} \frac{1-\prod_{i \in S} z_i}{\prod_{i \in S}(1-z_i)} \prod_{i \in S} \frac{\dd z_i}{2\pi \ii z_i}  z_i^{x-y_i}f^t(z_i), \label{nth_particle}
\end{align}
where the sum goes over all subsets $S \subset \{1, 2, \ldots, N\}$ with $k$ elements while $c_R$ are large positively-oriented simple contours that encircle all the singularities of the integrand and we denote
$f(z) = \frac{1 + (z-2)\Gamma}{z-\Gamma}$.
To proceed, we multiply Eq.~\eqref{nth_particle} with $\mu^n$ and sum over $n=1, 2, \ldots, N$ resulting in
\begin{align}
&\mathbb{E}_{Y}(\mu^{N_x}\eta_x|t)  \label{EY_weta} \\
&= \mu \sum_{n=1}^N(-\mu)^{n-1} \sum_{n\leq k\leq N}\binom{k-1}{n-1}    \sum_{|S|=k}\oint_{c_R}^{\times k} \prod_{\stackrel{i,j \in S}{i<j}} \frac{z_j-z_i}{1-2z_i + z_i z_j} \frac{1-\prod_{i \in S} z_i}{\prod_{i \in S}(1-z_i)} \prod_{i \in S} \frac{\dd z_i}{2\pi \ii z_i}  z_i^{x-y_i} f^t(z_i) \nonumber\\ 
&= \mu \sum_{k=1}^N \sum_{n=1}^{k}(-\mu)^{n-1} \binom{k-1}{n-1}    \sum_{|S|=k}\oint_{c_R}^{\times k}  \prod_{\stackrel{i,j \in S}{i<j}} \frac{z_j-z_i}{1-2z_i + z_i z_j}  \frac{1-\prod_{i \in S} z_i}{\prod_{i \in S}(1-z_i)} \prod_{i \in S} \frac{\dd z_i}{2\pi \ii z_i}  z_i^{x-y_i} f^t(z_i)  \nonumber \\ 
&= \mu \sum_{k=1}^N(1-\mu)^{k-1}   \sum_{|S|=k}\oint_{c_R}^{\times k} \prod_{\stackrel{i,j \in S}{i<j}} \frac{z_j-z_i}{1-2z_i + z_i z_j}  \frac{1-\prod_{i \in S} z_i}{\prod_{i \in S}(1-z_i)} \prod_{i \in S} \frac{\dd z_i}{2\pi \ii z_i}  z_i^{x-y_i} f^t(z_i) \nonumber \\
&= \frac{\mu}{1-\mu} \sum_{k=1}^N(1-\mu)^{k}   \sum_{|S|=k}\oint_{c_R}^{\times k} \prod_{\stackrel{i,j \in S}{i<j}} \frac{z_j-z_i}{1-2z_i + z_i z_j} \frac{1-\prod_{i \in S} z_i}{\prod_{i \in S}(1-z_i)} \prod_{i \in S} \frac{\dd z_i}{2\pi \ii z_i}  z_i^{x-y_i} f^t(z_i), \nonumber
\end{align}
where we have introduced an indicator function
\begin{equation}
\eta_x(X) =
\begin{cases}
1 & x_i = x\ \textrm{for some } i,\\
0 & x_i \neq x\  \textrm{for all } i.
\end{cases}
\end{equation}
We now consider the step initial condition $y_i=i$ and send $N \to \infty$ in Eq.~\eqref{EY_weta}
\begin{align}
&\mathbb{E}_{\rm step}(\mu^{N_x}\eta_x|t)  = \frac{\mu}{1-\mu} \sum_{k=1}^\infty(1-\mu)^{k}   \sum_{|S|=k} \nonumber \\
&\oint_{c_R}^{\times k} \prod_{\stackrel{i,j \in S}{i<j}} \frac{z_j-z_i}{1-2z_i + z_i z_j}  \frac{1-\prod_{i \in S} z_i}{\prod_{i \in S}(1-z_i)} \prod_{i \in S} \frac{\dd z_i}{2\pi \ii z_i}  z_i^{x-i} f^t(z_i)\label{Estep_weta}.
\end{align}
The sum over the subsets of $S = \{1, 2, \ldots\}$ of size $k$ is computed as
\begin{equation}
\sum_{|S|=k} \prod_{i \in S} z^{-i} = \sum_{1\leq S_1 < S_2 < \ldots S_k} \prod_{i=1}^k z_i^{-S_i} = \prod_{i=1}^k \frac{z_i^{-1} \ldots z_k^{-1}}{1- z_i^{-1} \ldots z_k^{-1}}= \prod_{i=1}^k \frac{1}{z_i \ldots z_k -1}.
\end{equation}
We then have
\begin{align}
&\mathbb{E}_{\rm step}(\mu^{N_x}\eta_x|t)  = \frac{\mu}{1-\mu} \sum_{k=1}^\infty(1-\mu)^{k} \nonumber\\
 & \oint_{c_R}^{\times k} \prod_{1\leq i , j \leq k} \frac{z_j-z_i}{1-2z_i + z_i z_j}  \frac{1-\prod_{i=1}^k z_i}{\prod_{i=1}^k(1-z_i)} \prod_{i=1}^k \frac{\dd z_i}{2\pi \ii z_i}  \frac{z_i^{x}}{z_i \ldots z_k -1}f^t(z_i)\label{Estep_weta2}.
\end{align}
Since the integrals over all $z_i$-contours are identical we can symmetrize the above expression by summing over all permutations of $z_i$ variables, see Eq.~\eqref{sym_def}. Using Eq.~\eqref{sym2_equation} to simplify Eq.~\eqref{Estep_weta2} we have 
\begin{align}
&\mathbb{E}_{\rm step}(\mu^{N_x}\eta_x|t)   = \frac{\mu}{1-\mu} \sum_{k=1}^\infty\frac{(\mu-1)^k}{k!}    \nonumber \\
 &\oint_{c_R}^{\times k}\prod_{1\leq i , j \leq k} \frac{z_j-z_i}{1-2z_i + z_i z_j}  \frac{1-\prod_{i=1}^k z_i}{\prod_{i=1}^k(1-z_i)^2} \prod_{i=1}^k \frac{\dd z_i}{2\pi \ii z_i}  z_i^{x} f^t(z_i) \label{Estep_weta3}.
\end{align}
To eliminate the $\eta_x$ term from the average we observe that
\begin{equation}
\mu^{N_x} = \mu^{N_x-1} + \mu^{N_x} \eta_x(1 - \mu^{-1}),
\end{equation}
which after telescoping leads to
\begin{equation}
\mu^{N_x}  = \lim_{y\to -\infty}\mu^{N_y}  + (1-\mu^{-1})\sum_{y=-\infty}^x \mu^{N_y}\eta_y  = 1 + (1-\mu^{-1})\sum_{y=-\infty}^x \mu^{N_y} \eta_y.
\end{equation}
Summing Eq.~\eqref{Estep_weta3} leaves us with
\begin{equation}
\mathbb{E}_{\rm step}(\mu^{N_x}|t)  =  \sum_{k=0}^\infty\frac{(\mu-1)^k}{k!}  \oint_{c_R}^{\times k}  \prod_{1\leq i < j \leq k} \frac{z_j-z_i}{1-2z_i + z_i z_j} \prod_{i=1}^k \frac{\dd z_i}{2\pi \ii}  \frac{z_i^{x}}{(1-z_i)^2} f^t(z_i)\label{Estep_weta4},
\end{equation}
where we have additionally absorbed the $1$ as the $k=0$ term. We now aim to bring the double product over all $i \neq j$ to be able to apply the determinant relation \eqref{TW_det_sum}. To this end, we rewrite Eq.~\eqref{Estep_weta4} as
\begin{align}
&\mathbb{E}_{\rm step}(\mu^{N_x}|t)   =  \sum_{k=0}^\infty\frac{(\mu-1)^k}{k!}  \oint_{c_R}^{\times k}  \prod_{1\leq i \neq j \leq k} \frac{z_j-z_i}{1-2z_i + z_i z_j} \nonumber \\
& \prod_{1\leq j < i \leq k} \frac{1-2z_i + z_i z_j} {z_j-z_i}\prod_{i=1}^k \frac{\dd z_i}{2\pi \ii}  \frac{z_i^{x}}{(1-z_i)^2} f^t(z_i)\label{Estep_weta5}.
\end{align}
Observe that all terms but the second product are symmetric under arbitrary permutations of variables. Using the factorization of symmetrization \eqref{sym_factor} and  the symmetrization relation \eqref{sym_eq_fac1} this factor is equal to unity
\begin{equation}
\mathbb{E}_{\rm step}(\mu^{N_x}|t)  =  \sum_{k=0}^\infty\frac{(\mu-1)^k}{k!}  \oint_{c_R}^{\times k} \prod_{1\leq i \neq j \leq k} \frac{z_j-z_i}{1-2z_i + z_i z_j} \prod_{i=1}^k \frac{\dd z_i}{2\pi \ii}  \frac{z_i^{x}}{(1-z_i)^2} f^t(z_i).
\label{Estep_weta6}
\end{equation}
Noting that the first product involves an even number of terms we can change the sign in the denominator, which leaves us in a position to use the determinant relation \eqref{TW_det_sum} and absorb the remaining $z_i$-dependence by using multilinearity of the determinant
\begin{equation}
\mathbb{E}_{\rm step}(\mu^{N_x}|t)  =  \sum_{k=0}^\infty\frac{(\mu-1)^k}{k!}  \oint_{c_R}^{\times k}\prod_{i=1}^k \frac{\dd z_i}{2\pi \ii}   \det \left( \frac{z_i^{-x}f^t(z_i) }{1 - 2z_i + z_i z_j } \right)_{1\leq i,j \leq k}\label{Estep_weta_final}.
\end{equation}
In the following we find it more convenient to work with small integration contours which we achieve by inverting the integration variables, $z_i \to 1/z_i$, resulting in
\begin{equation}
\mathbb{E}_{\rm step}(\mu^{N_x}|t)  =  \sum_{k=0}^\infty\frac{(\mu-1)^k}{k!}  \oint_{c_r}^{\times k} \prod_{i=1}^k \frac{\dd z_i}{2\pi \ii}   \det \left( \frac{z_i^{t-x}h^t(z_i) }{1 - 2z_i + z_i z_j } \right)_{1\leq i,j \leq k}\label{Estep_weta_final2},
\end{equation}
where $c_r$ is a small positively oriented circle that encircles only the singularity at the origin while the function $h$ is an exponentiated  `discrete dispersion relation'
\begin{equation}
h(z) = z^{-1} f(z^{-1}) = \frac{1 + (z^{-1} - 2)\Gamma}{1 - \Gamma z}.
\end{equation}
We note that by considering a square lattice with $x=t$ and taking the continuous-time limit $t \to \infty$ with $t \Gamma$ finite we find 
\begin{equation}
\lim_{\stackrel{\Gamma \to 0}{\Gamma t = T}} h^t(z) 
= e^{T(z+z^{-1}-2)},
\end{equation}
and Eq.~\eqref{Estep_weta_final} recovers the multiple integral representation of the simple symmetric exclusion process' full counting statistics obtained by Derrida and Gerschenfeld \cite{Derrida_Gerschenfeld_2009}. Eq.~\eqref{Estep_weta_final2} generalizes the analogous continuous-time result for the symmetric exclusion process, see Theorem 1.1 of Ref.~\cite{Imamura_Mallick_Sasamoto_2021}.\\
 
\noindent Combining the result \eqref{Estep_weta_final2} with the expression for the dressed FCS \eqref{dressed_6_vertex_partition_step}  and exchanging the order of summation and integration we can sum up the vacancy probabilities
\begin{align}
\langle e^{\lambda J(t)} \rangle = &\sum_{k=0}^\infty\frac{(\mu-1)^k}{k!}  \oint_{c_r}^{\times k} \prod_{i=1}^k \frac{\dd z_i}{2\pi \ii} \left[\overline \rho_-/\mu_- + \rho_-  \prod_{i=1}^k h(z_i)/z_i \right]^t \label{vacancy_FCS}\\
&\left[\overline \rho_+ \mu_- \prod_{i=1}^k z_i + \rho_+\right]^t \det \left( \frac{1}{1 - 2z_i + z_i z_j } \right)_{1\leq i,j \leq k} \nonumber .
\end{align}
It is presently not clear whether the expression \eqref{vacancy_FCS} can be written in a form that would facilitate its asymptotic analysis. We instead rewrite Eq.~\eqref{Estep_weta_final2} as a Fredholm determinant and extract its asymptotic behavior in Section \ref{sec:asymptotics} and only then return to the analysis of the dressed FCS \eqref{dressed_6_vertex_partition_step} in Section \ref{sec:dressed_asymptotics}.

\subsection{Fredholm determinant representation}
By introducing the integral kernel
\begin{equation}
K_{x, t}(z, z') = \frac{z^{t-x} h^t(z)}{1 - 2z + zz'} \label{K_def}
\end{equation}
that acts on functions as
\begin{equation}
K[g(z)] = \oint_{c}\frac{\dd z'}{2\pi \ii} K(z,z') g(z'),
\end{equation}
the expression \eqref{Estep_weta_final2} can be written as a Fredholm determinant
\begin{equation}
\mathbb{E}_{\rm step}(\mu^{N_x}|t) = {\rm det}_{{c}_r} [1 + (\mu-1)K_{x, t}],
\end{equation}
where the subscript denotes the integral contour of the kernel. Using the trace relation 
\begin{align}
&\log [{\rm det}_{{c}}(1+ v K)] = {\rm tr}_{{c}}[\log (1+vK)] \nonumber\\
= - &\sum^{\infty}_{k=1} \frac{(-v)^k}{k} \oint^{\times k}_{c} \prod_{i=1}^k \frac{\dd z_i}{2\pi \ii} K(z_1, z_2)\ldots K(z_k, z_1),
\end{align}
we obtain the following series representation of the full counting statistics
\begin{equation}
\log \mathbb{E}_{\rm step}(\mu^{N_x}|t) = - \sum^{\infty}_{k=1} \frac{(1-\mu)^k}{k} I_{x, t}^{(k)} \label{CGF_int}
\end{equation}
in terms of traces of powers of  the integral kernel \eqref{K_def}
\begin{equation}
I_{x, t}^{(k)} = {\rm tr}_{{c}_r}\, K^k_{x, t} = \oint^{\times k}_{\mathcal{C}_r} \prod_{i=1}^k \frac{\dd z_i}{2\pi \ii}  \frac{z_i^{t-x}h^t(z_i)}{1 - 2z_i + z_iz_{i+1}}. \label{trace_def}
\end{equation}
where we identify variables in the $k$-th term periodically, i.e.~$z_{k+1} \equiv z_1$.

\subsection{Alternative trace representation}
To facilitate the analysis of the traces $I_{x, t}^{(k)}$ \eqref{trace_def} we rewrite them in an alternative form inspired by Derrida and Gerschenfeld's treatment of the simple symmetric exclusion process in Ref.~\cite{Derrida_Gerschenfeld_2009}.
Noting that
\begin{equation}
\frac{1 + (z_i^{-1} - 2)\Gamma}{1-\Gamma z_{i+1}} = 1 + \frac{(z_i^{-1} - 2 + z_{i+1})\Gamma}{1-\Gamma z_{i+1}}
\end{equation}
we can rewrite the trace \eqref{trace_def} as
\begin{equation}
 I_{x, t}^{(k)} = \oint^{\times k}_{c_r} \prod_{i=1}^k \frac{\dd z_i}{2\pi \ii z_i}  z_i^{t-x} \frac{\left(1 + \frac{(z_i^{-1} - 2 + z_{i+1})\Gamma}{1-\Gamma z_{i+1}}\right)^t}{z_i^{-1}-2 + z_{i+1}}. \label{Ik_1}
\end{equation}
By summing a geometric series we have 
\begin{equation}
\frac{\left(1 + \frac{(z_i^{-1} - 2 + z_{i+1})\Gamma}{1-\Gamma z_{i+1}}\right)^t}{z_i^{-1}-2 + z_{i+1}}= \frac{1}{z_i^{-1}-2 + z_{i+1}} + \frac{\Gamma}{1 - \Gamma z_{i+1}} \sum_{t_i=0}^{t-1} \left(1 + \frac{(z_i^{-1} - 2 + z_{i+1})\Gamma}{1-\Gamma z_{i+1}}\right)^{t_i}
\end{equation}
so that 
\begin{align}
& I_{x, t}^{(k)}  = \sum_{E \subset \{1, 2, \ldots, k\}} \oint^{\times k}_{c_r}  \left[\prod_{i=1}^k \frac{\dd z_i}{2\pi \ii z_i}  z_i^{t-x}\right]\\
&\left[\prod_{i \notin E} \frac{1}{z_i^{-1}-2 + z_{i+1}}  \right] 
\left[\prod_{i \in E} \frac{\Gamma}{1 - \Gamma z_{i+1}} \sum_{t_i=0}^{t-1} \left(1 + \frac{(z_i^{-1} - 2 + z_{i+1})\Gamma}{1-\Gamma z_{i+1}}\right)^{t_i} \right]. \label{I_E_int}
\end{align}
However, if the set $E \neq \{1, 2, \ldots, k\}$ there exists a variable $z_i$ whose integral reads either
\begin{equation}
\oint_{c_r} \frac{\dd z_i}{2\pi \ii z_i}z_i^{t-x} \frac{\left(1 + \frac{(z_{i-1}^{-1} - 2 + z_{i})\Gamma}{1-\Gamma z_i}\right)^{t_{i-1}} }{(1 - \Gamma z_i)(z_i^{-1}-2 + z_{i+1})} 
\end{equation}
or
\begin{equation}
\oint_{c_r} \frac{\dd z_i}{2\pi \ii z_i} z_i^{t-x} \ \frac{1}{(z_{i-1}^{-1}-2 + z_{i})(z_i^{-1}-2 + z_{i+1})}.
\end{equation}
Both of these integrals vanish for $t \geq x$ and the only non-zero contribution to the integral \eqref{I_E_int} is from $E = \{1, 2, \ldots, k\}$
\begin{equation}
 I_{x, t}^{(k)}  = \Gamma^k \oint_{c_r}^{\times k}  \prod_{i=1}^k \sum_{t_i=0}^{t-1}   \frac{\dd z_i}{2\pi \ii z_i}  
 \frac{ z_i^{t-x}}{1 -\Gamma z_{i}}  \left(1 + \frac{(z_i^{-1} - 2 + z_{i+1})\Gamma}{1-\Gamma z_{i+1}}\right)^{t_i} \quad {\rm for}\ {t \geq x},
\end{equation}
which simplifies to
\begin{equation}
 I_{x, t}^{(k)}  =  \Gamma^k   \oint_{c_r}^{\times k} \prod_{i=1}^k \sum_{t_i=0}^{t-1}  \frac{\dd z_i}{2\pi \ii z_i}  
  z_i^{t-x} \frac{\left(1 + (z_i^{-1} - 2)\Gamma\right)^{t_i}}{\left(1-\Gamma z_{i}\right)^{t_{i-1}+1}} \quad {\rm for}\ {t \geq x}.
 \label{I_representation}
\end{equation}
The representation \eqref{I_representation} is the starting point for the asymptotic analysis of the traces performed in Section~\ref{sec:asymptotics}. While the representation \eqref{I_representation} is valid only for $t \geq x$, we can use a reflection relation to extend the result to the regime $t <x$.

\subsubsection{Reflection relation} Observing that the six-vertex matrix \eqref{six_vertex} is symmetric with respect to reflections over either diagonal, the corresponding vertex \eqref{V_vert_def} is invariant under a `particle-hole' transformation
\begin{equation}
U_{\, h\,\,  v}^{h' v'} = U_{\, \overline h\,\,  \overline v}^{\overline h' \overline v'} \label{particle_hole}
\end{equation}
where $\overline \bullet$ maps particles to holes and vice-versa, i.e.~$\overline \uparrow = \downarrow$ and $\overline \downarrow = \uparrow$. We also note that a rectangular lattice of the bistochastic six-vertex model is invariant under reflection over its diagonal
 \begin{equation}
\mathcal{M}_{h_{0, 1} v_{i,0}}^{h_{x, 1} v_{i,1}}\mathcal{M}_{h_{0, 2} v_{i,1}}^{h_{x, 2} v_{i,2}}\ldots \mathcal{M}_{h_{x, t} v_{i,t-1}}^{h_{x, t} v_{i,t}} =
\mathcal{M}_{v_{1, 0} h_{0,j}}^{v_{t, 0} h_{1,j}}\mathcal{M}_{v_{1, 1} h_{1,j}}^{v_{t, 1} h_{2,j}}\ldots \mathcal{M}_{v_{1, 0} h_{x-1,j}}^{v_{t, 0} h_{x,j}} \label{mirror}
\end{equation}
where the sub-indices  in the second index of each monodromy matrix take values $ 1 \leq i\leq x$ and $1 \leq j \leq t$ and indicate correspodning strings of symbols. Composing the particle hole-transformation \eqref{particle_hole} and mirroring the lattice using \eqref{mirror} and noting that this maps step initial conditions on one rectangular lattice to step initial condition on the mirrored rectangular lattice, we obtain a reflection identity for the full counting statistics \eqref{E_FCS_def} with step initial conditions
\begin{equation}
\mathbb{E}_{\rm step}(\mu^{N_x}|t)  \mu^t  =  \mathbb{E}_{\rm step}(\mu^{N_t}|x)  \mu^x. \label{step_reflection}
\end{equation}

\subsubsection{Free and single-file limits}
Starting from \eqref{I_representation}, it is instructive to consider the two limiting cases of $\Gamma = 0$ and $\Gamma =1$ which correspond to single-file and free dynamics of the cellular automaton respectively. 

\paragraph{Single-file limit} The integrand in \eqref{I_representation} is finite. For $\Gamma =0$ the traces trivially vanish for $t\geq x$ due to the vanishing $\Gamma^k$ prefactor 
\begin{equation}
I_{x, t}^{(k)}|_{\Gamma =0}  =  0 \quad {\rm for}\ {t \geq x}.
\end{equation}
The series \eqref{CGF_int} then immediately gives $\mathbb{E}_{\rm step}(\mu^{N_x}|t)|_{\Gamma =0} =1$ for $t \geq x$. The reflection relation \eqref{step_reflection} extends the result to $t < x$
\begin{equation}
\mathbb{E}_{\rm step}(\mu^{N_x}|t)|_{\Gamma =0} =
\begin{cases}
1 & {\rm for}\ t \geq x,\\
\mu^{x-t} & {\rm for}\ t < x.
\end{cases}
\end{equation}

\paragraph{Free limit}  For $\Gamma = 1$ the representation \eqref{I_representation} simplifies to
\begin{equation}
 I_{x, t}^{(k)}|_{\Gamma =1 } 
  = \oint_{c_r}^{\times k}  \sum_{t_i=0}^{t-1}    \frac{\dd z_i}{2\pi \ii z_i}  
 \frac{z_i^{t-x-t_i}}{1-z_i}  \left(\frac{1-z_i}{1-z_{i+1}}\right)^{t_i} \quad {\rm for}\ {x \leq t}.  \label{free_I_representation}
\end{equation}
Expanding the numerator and denominator using the binomial theorem we have
\begin{align}
 I_{x, t}^{(k)}|_{\Gamma =1 } 
 &= \oint_{c_r}^{\times k}  \prod_{i=1}^k  \sum_{t_i=0}^{t-1}    \frac{\dd z_i}{2\pi \ii z_i}  
 z_i^{t-x-t_i}  (1-z_i)^{t_i - t_{i-1}-1} \nonumber \\
 &= \oint_{c_r}^{\times k}  \prod_{i=1}^k  \sum_{t_i=0}^{t-1}\sum_{u_i=0}^\infty    \frac{\dd z_i}{2\pi \ii z_i}  
\binom{t_{i-1} - t_i + u_i}{u_i}  z_i^{t-x-t_i + u_i} \nonumber\\
 &= \prod_{i=1}^k  \sum_{t_i=0}^{t-1}\sum_{u_i=0}^\infty   
\binom{t_{i-1} - t_i + u_i}{u_i}  \delta_{t-x-t_i + u_i, 0} \nonumber\\
 &= \prod_{i=1}^k  \sum_{t_i=t-x}^{t-1}   
\binom{t_{i-1} + x -t}{t_i + x -t} = \prod_{i=1}^k  \sum_{s_i=0}^{x-1}   
\binom{s_{i-1}}{s_i} \nonumber \\
 &=  \sum_{s_1, s_2, \ldots, s_k=0}^{x-1}   
\binom{s_1}{s_2} \binom{s_2}{s_3} \ldots \binom{s_k}{s_1}. \label{free_sum}
\end{align}
Due to the periodic structure of the expression there exists an index $i$ such that $s_i \leq s_{i+1}$.
We now observe that the binomial coefficient with integer entries vanishes,
$\binom{n}{k} = 0$ for  $n < k$ with  $n, k \in \mathbb{N}$.
The only non-zero term in the sum over the variable $s_i$ satisfying $s_i \leq s_{i+1}$ is therefore the one with $s_i = s_{i+1}$,
 which yields an equivalent problem with one variable less. Repeating the same procedure $k-1$-times we come to
\begin{equation}
 I_{x, t}^{(k)}|_{\Gamma =1 }  =  \sum_{s_i=0}^{x-1} \label{free_trace_sum}   
\binom{s_i}{s_i} = x \quad {\rm for}\ t \geq x.
\end{equation}
The series \eqref{CGF_int} now sums to
\begin{equation}
\mathbb{E}_{\rm step}(\mu^{N_x}|t)|_{\Gamma =1} = \mu^x. \label{free_CGF}
\end{equation}
While the result \eqref{free_trace_sum} was derived for $t \geq x$, is is easy to see that \eqref{free_CGF} holds for all $x$ and $t$ using the reflection relation \eqref{step_reflection}.

\section{Asymptotics}
\label{sec:asymptotics}
We are now in a position to analyze the asymptotic behavior of the traces. Starting from the multiple-integral representation \eqref{I_representation} we evaluate the contour integrals, resulting in a double sum representation. We next obtain the asymptotic form of the function defined by the inner summation. The asymptotic form of the traces follows from the asymptotic analysis of the outer summation.

\subsection{Trace expansion}
 Guided by the computation   \eqref{free_sum}  in the free case and noting that $1 + (z^{-1} - 2)\Gamma = \frac{(2\Gamma -1)(1-\Gamma z) + \overline \Gamma^2}{\Gamma z}$
we use the binomial theorem $(1+z)^{n \geq 0} = \sum_{j=0}^\infty \binom{n}{j}z^j$ and the generalized binomial theorem $(1-z)^{-n  < 0} = \sum_{j=0}^\infty \binom{n+j-1}{n-1}z^j $ for   $|z|< 1$
to expand the representation \eqref{I_representation} for $t \geq x$ as
\begin{align}
 &I_{x, t}^{(k)} =   \Gamma^k   \oint_{c_r}^{\times k} \prod_{i=1}^k \sum_{t_i=0}^{t-1}  \frac{\dd z_i}{2\pi \ii z_i}  
  z_i^{t-x-t_i} \Gamma^{-t_i} \frac{\left((2\Gamma -1)(1-\Gamma z) + \overline \Gamma^2\right)^{t_i}}{\left(1-\Gamma z_{i}\right)^{t_{i-1}+1}}\\
&=  \Gamma^k   \oint_{c_r}^{\times k} \prod_{i=1}^k \sum_{t_i=0}^{t-1} \sum_{p_i=0}^{t_i} \binom{t_i}{p_i}   \frac{\dd z_i}{2\pi \ii z_i}   \Gamma^{-t_i} \overline \Gamma^{2p_i} (2\Gamma -1)^{t_i-p_i} \frac{z_i^{t-x-t_i}}{(1-\Gamma z_i)^{t_{i-1}-t_i +p_i +1}} \nonumber\\
&=  \Gamma^k   \oint_{c_r}^{\times k} \prod_{i=1}^k \sum_{t_i=0}^{t-1} \sum_{p_i=0}^{t_i} \sum_{u_i=0}^{\infty} \binom{t_i}{p_i} \binom{t_{i-1} - t_i + p_i + u_i}{u_i} \frac{\dd z_i}{2\pi \ii z_i}   \Gamma^{u_i-t_i} \overline \Gamma^{2p_i} (2\Gamma -1)^{t_i-p_i} z_i^{t-x-t_i+u_i} \nonumber\\
&=  \Gamma^k   \prod_{i=1}^k \sum_{t_i=0}^{t-1} \sum_{p_i=0}^{t_i} \sum_{u_i=0}^{\infty} \binom{t_i}{p_i} \binom{t_{i-1} - t_i + p_i + u_i}{u_i}   \Gamma^{u_i-t_i} \overline \Gamma^{2p_i} (2\Gamma -1)^{t_i-p_i} \delta_{t-x-t_i+u_i, 0}. \nonumber
\end{align}
Before summing over the $\delta$-function we use the Chu-Vandermond identity, $\binom{n+m}{k} = \sum_{j=0}^k \binom{n}{j} \binom{m}{k-j}$, 
to extract and resum the $p_i$ term in the second binomial
\begin{align}
 &I_{x, t}^{(k)}=  \Gamma^k   \prod_{i=1}^k \sum_{t_i=0}^{t-1} \sum_{p_i=0}^{t_i} \sum_{u_i=0}^{\infty} \sum_{j_i=0}^{u_i} \binom{t_i}{p_i} \binom{p_i}{j_i} \binom{t_{i-1} - t_i  + u_i}{u_i-j_i}   \Gamma^{u_i-t_i} \overline \Gamma^{2p_i} (2\Gamma -1)^{t_i-p_i} \delta_{t-x-t_i+u_i, 0} \nonumber\\
&=  \Gamma^k   \prod_{i=1}^k \sum_{t_i=0}^{t-1} \sum_{p_i=0}^{t_i} \sum_{u_i=0}^{\infty} \sum_{j_i=0}^{u_i} \binom{t_i}{j_i} \binom{t_i-j_i}{t_i-p_i} \binom{t_{i-1} - t_i  + u_i}{u_i-j_i}   \Gamma^{u_i-t_i} \overline \Gamma^{2p_i} (2\Gamma -1)^{t_i-p_i} \delta_{t-x-t_i+u_i, 0} \nonumber\\
&=  \Gamma^k   \prod_{i=1}^k \sum_{t_i=0}^{t-1}  \sum_{u_i=0}^{\infty} \sum_{j_i=0}^{u_i} \binom{t_i}{j_i} \binom{t_{i-1} - t_i  + u_i}{u_i-j_i}   \Gamma^{u_i-t_i} \overline \Gamma^{2j_i}  \delta_{t-x-t_i+u_i, 0} \nonumber \\
& \qquad \sum_{p_i=j_i}^{t_i} \binom{t_i-j_i}{t_i-p_i} (2\Gamma -1)^{t_i-p_i}  \overline \Gamma^{2(p_i-j_i)}  \nonumber
\end{align}
\begin{align}
&=  \Gamma^k   \prod_{i=1}^k \sum_{t_i=0}^{t-1}  \sum_{u_i=0}^{\infty}  \Gamma^{u_i+ t_i}  \delta_{t-x-t_i+u_i, 0} \sum_{j_i=0}^{u_i} \binom{t_i}{j_i} \binom{t_{i-1} - t_i  + u_i}{u_i-j_i}   \left(\overline \Gamma/\Gamma\right)^{2j_i}  \nonumber\\
&=  \Gamma^{k}   \prod_{i=1}^k \sum_{t_i=t-x}^{t-1}  \Gamma^{2t_i + x -t}  \sum_{j_i=0}^{t_i+x-t} \binom{t_i}{j_i} \binom{t_{i-1}  + x - t}{t_i + x - t-j_i}   \left(\overline \Gamma/\Gamma\right)^{2j_i}  \nonumber\\
&=  \Gamma^{k}   \prod_{i=1}^k \sum_{s_i=0}^{x-1}  \Gamma^{2 s_i + t-x}  \sum_{j_i=0}^{s_i} \binom{s_i + t-x}{j_i} \binom{s_{i-1}}{s_i -j_i}   \left(\overline \Gamma/\Gamma\right)^{2j_i} \nonumber \\
&= \Gamma^k \sum_{s_1, \ldots, s_k=0}^{x-1} \psi^{(k)}_{t- x}(s_1, \ldots, s_k; \gamma) \label{pseudo_Vandermond_representation}
\end{align}
where $\gamma = \overline \Gamma/\Gamma \geq 0$ and we have introduced the functions $\psi_n^{(k)}$
\begin{equation}
\psi_n^{(k)}(s_1, s_2, \ldots, s_k; \gamma) =  \prod_{i=1}^k  (1+\gamma)^{ - 2s_i-n}  \sum_{j_i=0}^{s_i} \binom{s_i +n}{j_i} \binom{s_{i-1}}{s_i -j_i}   \gamma^{2j_i}.
 \label{psi_def}
\end{equation}
The product of binomials in Eq.~\eqref{psi_def} can be expressed as a product of contour integrals
\begin{equation}
\psi_n^{(k)}(s_1, \ldots, s_k; \gamma) =      \prod_{i=1}^k  (1+\gamma)^{- 2s_i-n}  \oint_{c_r}\frac{\dd z_i}{2\pi \ii z_i}  z_i^{-s_i} (1 + \gamma^2 z_i)^{s_i + n}(1+z_i)^{s_{i-1}}, \label{alternative_integral_representation}
\end{equation}
which makes it manifest that $\psi_n^{(k)}$ are invariant under all permutations $\sigma \in \mathfrak{S}_k$ of  $\{s_i\}_{i=1}^k$ 
\begin{equation}
\psi_n^{(k)}(s_1, s_2, \ldots, s_k; \gamma)= \psi^{(k)}_{n}(s_{\sigma(1)}, s_{\sigma(2)}, \ldots, s_{\sigma(k)}; \gamma). \label{psi_sym}
\end{equation}
Expanding the binomials in Eq.~\eqref{psi_def} and using the series definition of the hypergeometric function
\be
{}_{2}F_1(a, b, c; z)= \sum_{n=0}^\infty \frac{(a)_n(b)_n}{(c)_n} \frac{z^n}{n},
\ee
where $(a)_n = a(a+1)\ldots (a+n-1)$ is the rising Pochhammer symbol, we find yet another representation of $\psi_n^{(k)}$
\begin{equation}
\psi_n^{(k)}(s_1, \ldots, s_k; \gamma) =  \prod_{i=1}^k (1+\gamma)^{- 2s_i-n}  \binom{s_{i-1}}{s_i} {}_{2}F_1(-n-s_i, -s_i, 1+ s_{i-1}-s_i,  \gamma^2). \label{psi_hypergeometric}
\end{equation}
An advantage of the representation \eqref{psi_hypergeometric} is that it immediately recovers the free result \eqref{free_trace_sum} for $\gamma=0$ since ${}_{2}F_1(a, b, c; 0) = 1$ for all $a, b, c \in \mathbb{R}$ leaving only the product of binomials. While the same product of binomials occurs for all values of $\gamma$, the function $\psi^{(k)}_n$ is not identically zero away from the diagonals ${s_1 = \ldots = s_k}$ since for $\gamma>0$, the hypergeometric function diverges for non-positive integer third arguments  and this counterbalances the vanishing of the binomials. In that case Eq.~\eqref{psi_hypergeometric} should be understood as the appropriate limit using the relation 
\be
\lim_{c \to -m} \frac{{}_2F_1(a, b, c; z)}{\Gamma(c)} = \frac{(a)_{m+1}(b)_{m+1}}{(m+1)!} z^{m+1}{}_2F_1(a+m+1, b+m+1, m+2; z),
\ee 
for $m\in \mathbb{Z}_{\geq 0}$. The maximum of $\psi^{(k)}_n$, however, still occurs on the diagonal by virtue of the permutation symmetry \eqref{psi_sym}. This will be important in the upcoming asymptotic analysis.

\subsection{Asymptotic form of $\psi^{(k)}_n$}
We now obtain the asymptotics of $\psi^{(k)}_n$ when $s_i$ and $n$ are large and both $\mathcal{O}(t^{\alpha})$ with $0 < \alpha \leq 1$. We accordingly introduce the rescaled variables
\begin{equation}
j_i = y_i t^{\alpha} \geq 0, \qquad s_i = \eta_i t^{\alpha} \geq 0, \qquad n = \delta t^{\alpha} \geq 0, \label{rescaled_vars}
\end{equation}
in terms of which we have for $t \to \infty$
\begin{equation}
\psi_{n = \delta t^{\alpha}}^{(k)}(s_1 = \eta_1 t^{\alpha}, \ldots, s_k =  \eta_k t^{\alpha}; \gamma)  \simeq \frac{1}{(2\pi)^{k/2}} \int_0^{\eta_1} \ldots \int_0^{\eta_k} \dd y_1 \ldots \dd y_k\ \vartheta_{n}^{(k)}e^{t^{\alpha}\varphi_{n}^{(k)}},
\end{equation}
with
\begin{equation}
\vartheta_{n}^{(k)} =  \prod_{i=1}^k \theta_\delta(\eta_{i-1}, \eta_i, y_i)\,.
\end{equation}
Here we set 
\begin{equation}
\theta_\delta(\eta_{i-1}, \eta_i, y_i) = \sqrt{\frac{(\eta_i + \delta)\eta_{i-1}}{2\pi y_i(\eta_i+\delta -y_i)(\eta_i-y_i)(\eta_{i-1} -\eta_i +y_i)}},
\end{equation}
while
\begin{equation}
\varphi_{n}^{(k)}= \sum_{i=1}^k  \phi_\delta(\eta_{i-1}, \eta_i, y_i),
\end{equation}
and 
\begin{align}
\phi_\delta&(\eta_{i-1}, \eta_i, y_i) = 2y_i \log \gamma - (2\eta_i+\delta) \log (1+\gamma) \label{phi_explicit_def}\\
&+ (\eta_i+\delta)\log (\eta_i+\delta) - y_i \log y_i  - (\eta_i+ \delta- y_i) \log (\eta_i+ \delta- y_i) \nonumber \\
&+ \eta_{i} \log(\eta_{i}) - (\eta_i-y_i)\log (\eta_i-y_i) -(\eta_{i-1} - \eta_i +y_i)\log(\eta_{i-1} - \eta_i +y_i). \nonumber
\end{align}
Noting that the functions $\phi_\delta$ and $\theta_\delta$ depend on only a single $y_i$, the integral factorizes
\begin{equation}
\psi_{n}^{(k)} \simeq \frac{1}{(2\pi)^{k/2}} \prod_{i=1}^k \int_0^{\eta_i} \dd y_i\, \theta_\delta(\eta_{i-1}, \eta_i, y_i) e^{t^{\alpha}\phi_{\delta}(\eta_{i-1}, \eta_i, y_i)}. \label{psi_factorized_int}
\end{equation}
For ${t \to \infty}$ each integral is then given by Laplace's method, which quantifies the localization of the integrals around the critical points $y_i^*$ of the function in the exponent
\begin{equation}
\partial_{y_i} \phi_\delta(\eta_{i-1}, \eta_i, y_i^*)  = 0. \label{multi_y_cond}
\end{equation}
The critical points satisfy the quadratic equation
\begin{equation}
(\eta_i+\delta -y_i^*)(\eta_i-y_i^*) \gamma^2 = y_i^*(\eta_{i-1} - \eta_i + y_i^*) \label{y_quadratic}
\end{equation}
with the solutions 
\begin{equation}
y_i^{\pm}(\eta_{i-1}, \eta_{i}) = \frac{\gamma^2(2\eta_i+\delta) - \eta_i + \eta_{i-1} \pm \sqrt{D(\eta_{i-1}, \eta_i)}}{2(\gamma^2-1)}, \label{y_expression}
\end{equation}
where the discriminant reads $D(\eta_{i-1}, \eta_i) =(\gamma^2(2\eta_i+\delta) -\eta_i + \eta_{i-1})^2 - 4 \gamma^2(\gamma^2-1)\eta_i(\eta_i + \delta)$.
A direct calculation establishes the bounds
\begin{equation}
|y_i^+| > \eta_i , \qquad \eta_i > y_i^- > 0, \label{y_bounds}
\end{equation}
so that only $y_i^-$ lies in the domain of integration in \eqref{psi_factorized_int}.
We also have
\begin{equation}
\partial^2_{y_i} \phi_\delta(\eta_{i-1}, \eta_i, y_i)  = -\left[ \frac{1}{y_i} + \frac{1}{\eta_i + \delta - y_i} + \frac{1}{\eta_i -y_i} + \frac{1}{\eta_{i-1}-\eta_i + y_i} \right].
\end{equation}
Using \eqref{y_quadratic} and \eqref{y_bounds} it is straightforward to show that
\begin{equation} 
\partial^2_{y_i} \phi_\delta(\eta_{i-1}, \eta_i, y_i^-) < 0,
\end{equation}
indicting that $y_i^-=y_i^*$ is a local maximum around which the integrals \eqref{psi_factorized_int} localize. Denoting functions at the critical point $y_i^*$ as
\begin{align}
\Phi_\delta(\eta_{i-1}, \eta_i) &= \phi_\delta(\eta_{i-1}, \eta_i, y^*_i(\eta_{i-1}, \eta_i)), \label{f1}\\
\Phi_{\delta, 2}(\eta_{i-1}, \eta_i) &= \partial_{y_i}^2 \phi_\delta(\eta_{i-1}, \eta_i, y^*_i(\eta_{i-1}, \eta_i)), \label{f2} \\
\Theta_\delta(\eta_{i-1}, \eta_i) &= \theta_\delta(\eta_{i-1}, \eta_i, y^*_i(\eta_{i-1}, \eta_i)), \label{f3}
\end{align}
and localizing the integrals \eqref{psi_factorized_int} we now obtain the asymptotic form of $\psi^{(k)}_n$
\begin{equation}
\psi_{\delta t^{\alpha}}^{(k)}(\eta_1 t^{\alpha}, \ldots, \eta_k t^{\alpha}; \gamma) \simeq t^{-\alpha k/2} \Psi^{(k)}_{\delta; \alpha}(\eta_1, \ldots, \eta_k; \gamma), \label{psi_localized}
\end{equation}
where
\begin{equation}
\Psi^{(k)}_{\delta; \alpha}(\eta_1, \ldots, \eta_k; \gamma) = \prod_{i=1}^k \frac{\Theta_\delta(\eta_{i-1}, \eta_i) }{\sqrt{\Phi_{\delta, 2}(\eta_{i-1}, \eta_i)}} e^{t^{\alpha} \Phi_\delta(\eta_{i-1}, \eta_i)}. 
\end{equation}

\subsection{Trace asymptotics}
We now obtain the asymptotic forms of the traces for deviations from a square lattice on both diffusive and ballistic scales. Namely, we consider 
\begin{itemize}
\item[(i)] $t-x=\delta \sqrt{t}$;
\item[(ii)] $t-x=\delta {t}$;
\end{itemize}
with $\delta > 0$.  Expecting diffusive behavior of the stochastic six-vertex model motivates the diffusive scaling (i) while we also consider the ballistic scaling (ii) to facilitate the analysis of large fluctuations of the stochastic cellular automaton in Section \ref{sec:dressed_large_asymptotics}.

\subsubsection{Diffusive trace asymtptotics}
Converting the sum in Eq.~\eqref{pseudo_Vandermond_representation} into an integral and using the asymptotic form of \eqref{psi_localized} with $\alpha=1/2$ we have
\begin{equation}
I^{(k)}_{x, t} \simeq \Gamma^{k} t^{k/4}  \int_{0}^{x/t^{1/2}} \dd \eta_1\ldots \dd \eta_k\ \Psi^{(k)}_{\delta; 1/2}(\eta_1, \ldots, \eta_k; \gamma)\,, \qquad t-x = \delta t^{1/2} \geq 0.\label{trace_localization2}
\end{equation}
The asymptotic form of the integral \eqref{trace_localization2} is dictated by the maximum of $\Psi^{(k)}_\delta$ which occurs on the diagonal. 
To analyze the behavior along the diagonal we first control the divergent upper limit of the integral \eqref{trace_localization2} by rescaling variables as
\begin{equation}
\eta_i = \xi_i t^{1/2}. \label{eta_xi}
\end{equation}
The function $\Phi_\delta$ in the exponent has the following expansion in powers of $t^{1/2}$
\begin{equation}
\Phi_\delta(\xi_{i-1}t^{1/2}, \xi_{i}t^{1/2}) = t^{1/2}\sum_{p=0}^\infty \Phi_\delta^{[p]}(\xi_{i-1}, \xi_i) t^{-p/2}  \label{Phi_expansion}
\end{equation}
To obtain the local behavior around the diagonal we introduce the local coordinates
\begin{equation}
\xi_i = \xi + \epsilon \zeta_i \label{xi_zeta}
\end{equation}
and expand the leading order in the $t$-expansion \eqref{Phi_expansion} to second order in $\epsilon$
\begin{equation}
\Phi_\delta^{[0]}(\xi+\epsilon \zeta_{i-1}, \xi + \epsilon \zeta_i) = -(\zeta_{i-1} - \zeta_i) \log \frac{\gamma}{1+\gamma} \epsilon  - \frac{1}{4\gamma \xi} (\zeta_{i-1}-\zeta_i)^2 \epsilon^2 + \mathcal{O}(\epsilon^3).
\end{equation}
The leading time behavior of the function in the exponent near the diagonal is therefore
\begin{equation}
\sum_{i=1}^k \Phi_\delta(\xi+\epsilon \zeta_{i-1}, \xi + \epsilon \zeta_i) = -\frac{1}{4 \gamma \xi} \sum_{i=1}^k (\zeta_{i-1}-\zeta_i)^2 + \mathcal{O}(\zeta_i^3), \label{ridge_quadratic}
\end{equation}
where all $\zeta_i$ are of the same order. Eq.~\eqref{ridge_quadratic} shows that the exponent decays in directions transversal to the diagonal and the corresponding $(k-1)$-dimensional integral can be evaluated using Laplace's method. This leaves a function along the diagonal that can be extracted from the sub-leading order in the expansion \eqref{Phi_expansion}. The sum of two leading orders cancels on the diagonal
\begin{equation}
\sum_{i=1}^k \Phi_\delta^{[0]}(\xi, \xi) = \sum_{i=1}^k \Phi_\delta^{[1]}(\xi, \xi) = 0, 
\end{equation}
while the second subleading order is non-zero
\begin{equation}
\sum_{i=1}^k \Phi_\delta^{[2]}(\xi, \xi) = -\frac{k \gamma \delta^2}{4 \xi}.
\end{equation}
 In view of the above changes of variables \eqref{eta_xi} and \eqref{xi_zeta} the integral \eqref{trace_localization2} becomes
\begin{equation}
I^{(k)}_{x, t}  \simeq \Gamma^{k} t^{k/2} \int \dd \zeta_1\ldots \dd \zeta_k\  \frac{\Theta_{\delta; k}}{\sqrt{\Phi_{2,\delta;k}}} e^{-\frac{t^{1/2}}{4 \gamma \xi} \sum_{i=1}^k (\zeta_{i-1}-\zeta_i)^2}e^{-\frac{k \gamma \delta^2}{4 \xi}} , \label{trace_localization3}
\end{equation}
where, anticipating localization in transversal directions, we have suppressed integration boundaries and used Eqs.~\eqref{f2} and \eqref{f3} to expand the sub-exponential terms along the diagonal to leading order
\begin{equation}
\Theta_{\delta; k} = \left(\frac{(1+\gamma)^2}{\sqrt{2\pi} \gamma \xi} \right)^k, \qquad
\Phi_{2,\delta; k} = \left(-\frac{2(1+\gamma)^2}{\gamma\xi} \right)^k .
\end{equation}
To perform the integral \eqref{trace_localization3} we now make the following change of variables
\begin{equation}
(u, \nu_2, \ldots, \nu_{k})^T = R^{(k)} (\zeta_1, \zeta_2, \ldots, \zeta_k)^T, \label{zeta_nu}
\end{equation}
where the matrix $R^{(k)}$ reads
\begin{equation}
R^{(k)} =
\begin{bmatrix}
1 & 1 & 1 &  1 &\ldots \\
-1 & 1 & 0 & 0 & \ldots\\
0 & -1 & 1 & 0  &\ldots\\
0 & 0 & -1 & 1  &\ldots\\
\vdots & \vdots & \vdots & \vdots &\ddots
\end{bmatrix}.
\end{equation}
It is easy to establish a recursion for its determinant $\det R^{(k)} = 1 + \det R^{(k-1)}$ 
with the initial condition $R^{(1)}= 1$ and the solution $
 \det R^{(k)} = k.$
We therefore have a `diagonal' coordinate $
u = \sum_{i=1}^k \xi_k$
and $k-1$ `transversal' coordinates $
\nu_i = \xi_i -\xi_{i-1}.$
Note that the `cylic' difference is given by the sum of transversal coordinates $\xi_1 -\xi_k + = -\sum_{i=2}^k \nu_k.$
In terms of the rotated variables \eqref{zeta_nu} the integral \eqref{trace_localization3} reads
\begin{equation}
I^{(k)}_{x, t}  \simeq \Gamma^{k} t^{k/2} k^{-1} \int \dd u\dd \nu_2\ldots \dd \nu_k\  \frac{\Theta_{\delta; k}}{\sqrt{\Phi_{2,\delta;k}}} e^{-\frac{t^{1/2}}{2 \gamma \xi} \sum_{i, j=2}^k \nu_i \nu_j}e^{-\frac{k  \gamma\delta^2}{4\xi}}.
\label{trace_localization4}
\end{equation}
The Hessian matrix of the exponential term $-\frac{1}{2 \gamma \xi} \sum_{i, j=2}^k \nu_i \nu_j$ in the $(k-1)$-dimensional transversal subspace spanned by transversal coordiantes $\nu_{k\geq i\geq 2}$ reads
\begin{equation}
\left(H_\perp\right)_{i, j} = -\frac{1+\delta_{i,j}}{2 \gamma \xi}
\end{equation}
with the determinant $
{\rm det}\, H_\perp = \frac{k}{(-2 \gamma \xi)^{k-1}}$.
Applying Laplace's method to the integral over the  $k-1$ transversal directions we find
\begin{equation}
I^{(k)}_{x, t}  \simeq \Gamma^{k} (2\pi)^{(k-1)/2} t^{1/2} k^{-3/2} \int \dd u\ \xi^{(k-1)/2}  \frac{\Theta_{\delta; k}}{\sqrt{|{\rm det}\, H_\perp |\Phi_{2,\delta;k}}} e^{-\frac{k \gamma \delta^2}{4\xi} } \label{trace_localization5}.
\end{equation}
It remains to note that on the diagonal we have $
\dd u = k \dd \xi$ so that
collecting all the terms, simplifying and squaring the integration variable now brings us to the integral
\begin{equation}
I^{(k)}_{x, t} \simeq \sqrt{\frac{t }{\pi \gamma k}} \int_0^1 \dd \xi\, e^{-\frac{k \gamma \delta^2}{4\xi^2} }, \qquad \ t-x = \delta t^{1/2} \geq 0,
\label{trace_localization6}
\end{equation}
where the upper boundary is equal to unity due to $t - x = \mathcal{O}(t^{1/2})$.
The integral can be evaluated for $ k \gamma \delta^2>0$, giving the final result
\begin{equation}
I^{(k)}_{x, t} \simeq \sqrt{\frac{t}{\pi \gamma k}}\ p \left(\delta \sqrt{\gamma k/4}\right),\qquad\ t-x = \delta t^{1/2} \geq 0 
\label{trace_localization7},
\end{equation}
where we have introduced the function
\begin{equation}
p(z) = \int_0^1 \dd \xi\, e^{-z^2/\xi^2}=  e^{-z^2}  - z\sqrt{\pi}\, {\rm erfc} z\,, \label{p_def}
\end{equation}
with ${\rm erfc}\, z = \frac{2}{\sqrt{\pi}} \int_{z}^\infty \dd v\, e^{-v^2}$.
A consistency check of the final trace asymptotics \eqref{trace_localization7} is obtained by setting $\delta =0$ and taking the continuous-time limit
\begin{equation}
•I^{(k)}_{t, t} \simeq \sqrt{\frac{t}{\pi \gamma k}} \xrightarrow{\Gamma \to 0,\ \Gamma t = T}   \sqrt{\frac{T}{\pi k}}
\end{equation}
which recovers the result of Derrida and Gerschenfeld for continuous-time SSEP, see Appendix B of Ref.~\cite{Derrida_Gerschenfeld_2009}. We note that the trace asymptotics \eqref{trace_localization7} have also been obtained for the symmetric simple exclusion process, see Proposition 1.6 of Ref.~\cite{Imamura_Mallick_Sasamoto_2021}.

\subsubsection{Ballistic trace asymtptotics}
We again convert the sum in Eq.~\eqref{pseudo_Vandermond_representation} into an integral but this time using the asymptotic form of \eqref{psi_localized} with $\alpha=1$ to find
\begin{equation}
I^{(k)}_{x, t} \simeq \Gamma^{k} t^{k/2}  \int_{0}^{1-\delta} \dd \eta_1\ldots \dd \eta_k\ \Psi^{(k)}_{\delta; 1}(\eta_1, \ldots, \eta_k; \gamma), \qquad t-x = \delta t \geq 0.
\label{trace_localization2bal}
\end{equation}
As already noted, the maximum of the exponential part of $\Psi^{(k)}_{\delta; 1}$ occurs on the diagonal. To finds its location we compute the derivative of the exponential part on the diagonal
\begin{equation}
\partial_\eta \Phi_\delta(\eta, \eta) = \log \eta + \log(\eta + \delta) + 2\log \gamma -2 \log y^- - 2\log(1+\gamma) > 0,
\end{equation}
where $y^- = y^-(\eta, \eta)$ and the lower bound holds for $\eta > 0$ from \eqref{y_expression}. The maximum is therefore a boundary maximum on the end of the diagonal at $\eta_1 = \ldots= \eta_k = 1-\delta$. Computing the localization of the integral \eqref{trace_localization2} we have the following asymptotic form (up to sub-exponential terms)
\begin{equation}
   I^{(k)}_{x, t} \asymp e^{tk \Phi^b(\delta)} \quad {\rm for}\ t-x = \delta t \geq 0. \label{ballistic_trace} 
\end{equation}
where $ \Phi^b(\delta) = \Phi_{\delta}(1-\delta, 1-\delta)$.
Using Jensen's inequality it is easy to show that the exponential part defined in Eqs.~\eqref{phi_explicit_def} and \eqref{f1} is negative, $\Phi^b(\delta) < 0$ so that the traces vanish exponentially in $t$.

\subsection{Asymptotic form of the full counting statistics}
\label{sec:asymp_FCS}
Having obtained the asymptotic form of the traces we return to the series expansion of the full counting statistics
\eqref{CGF_int}. Using the integral representation \eqref{trace_localization6} and interchanging the order of summation and integration we have that for $t-x = \delta t^{1/2} \geq0$ the result reads 
\begin{equation}
\log \mathbb{E}_{\rm step}(\mu^{N_x}|t) \simeq -  \sqrt{\frac{t}{\pi \gamma}} \int_{0}^1 \dd \xi\, \sum^{\infty}_{k=1} k^{-3/2}
\left[(1-\mu)e^{-\frac{\gamma \delta^2}{4 \xi^2}} \right]^k. 
\label{CGF_sum}
\end{equation}
We note that equivalent asymptotics have been obtained for the symmetric simple exclusion process, see Theorem 1.5 of Ref.~\cite{Imamura_Mallick_Sasamoto_2021}.
The sum \eqref{CGF_sum} can be cast in an integral form using the identity 
\begin{equation}
-\sum_{k=1}^\infty \frac{(-z)^k}{k^{3/2}} = \frac{1}{\sqrt{\pi}}\int_{-\infty}^{\infty} \log \left(1+ z e^{-k^2} \right), 
\label{sum_to_integral}
\end{equation}
which follows by expanding the logarithm and integrating term by term. Substituting the identity \eqref{sum_to_integral} in the sum \eqref{CGF_sum} gives the final asymptotic form of the full counting statistics for $t-x = \delta t^{1/2} \geq0$. Explicitly we have
\begin{equation}
\log \mathbb{E}_{\rm step}(\mu^{N_x}|t) \simeq \sqrt{t\gamma} \int_{0}^1\dd \xi\, \int_{-\infty}^{\infty}\frac{\dd k}{\pi}\, \log \left(1+(\mu-1) e^{-k^2-\frac{\gamma \delta^2}{4 \xi^2} } \right).
\label{CGF_series_result}
\end{equation}

\noindent We also obtain the asymptotic form of the full-counting statistics for ballistic shape fluctuations. Since higher traces are exponentially suppressed it suffices to retain only the first (exponentially small) term in the series \eqref{CGF_int}, yielding
\begin{equation}
    \log \left[\log \mathbb{E}_{\rm step}(\mu^{N_x}|t) \right] \simeq  t \Phi^b(\delta), \qquad t-x = \delta t \geq 0. \label{CGF_balistic_asymptotics}
\end{equation}

\section{Equilibrium fluctuations of the stochastic charged cellular automaton}
\label{sec:dressed_asymptotics}
Having obtained the asymptotic form of the full counting statistics of the stochastic six-vertex model, we are in a position to consider the asymptotics off the dressed full counting statistics in Eq.~\eqref{dressed_6_vertex_partition_step}. While the expression \eqref{dressed_6_vertex_partition_step} allows for the analysis of any bipartite initial probability measure --- accomplished in Ref.~\cite{Krajnik2024} for the single-file case --- we presently restrict ourselves to equilibrium measures by setting (cf.~Eq.~\eqref{rho_pm})
\begin{equation}
\rho_\pm = \rho, \qquad b_\pm = b,
\end{equation}
where we have $\mu_\pm = \cosh \lambda \mp b \sinh \lambda$. With these choices, the dressed full counting statistics reads as 
\begin{equation}
\langle e^{\lambda J(t)} \rangle = \rho^{2t} \sum_{n_-, n_+ =0}^t \binom{t}{n_+} \binom{t}{n_-}\nu^{n_- + n_+}  \mu_-^{n_+ - n_-}   \mathbb{E}_{\rm step}(\mu^{N_{t-n_+}}|t-n_-),
 \label{dressed_6_vertex_partition_step_eq}
\end{equation}
where we set $\nu = \overline \rho/\rho$. Equilibrium charge fluctuations in the single-file limit of the stochastic cellular automaton were derived microscopically in \cite{Krajnik2022} and hydrodynamically in Ref.~\cite{Yoshimura2024}.

\subsection{Typical fluctuations}
The distribution of typical fluctuations on the scale $z$ (set by the variance) is defined as
\begin{equation}
P_{\rm typ}(j) = \lim_{t\to \infty} t^{1/2z} P(J=jt^{1/2z}|t).
\end{equation}
Recall that the FCS is the Laplace transform of the finite-time current distribution, see Eq.~\eqref{FCS_def}.
Dynamicaly rescaling the counting field as
\begin{equation}
 \lambda \to \lambda t^{-1/2z},
\label{dynamical_rescaling}
\end{equation}
 and considering $t \to \infty$ we have 
\be
\lim_{t\to \infty}  \langle e^{\lambda t^{-1/2z} J(t)} \rangle = \int \dd j P_{\rm typ}(j) e^{\lambda j}.
\ee 
Therefore, we can express the typical distribution in terms of the dynamically rescaled full counting statistics by inverting the Laplace transform
\begin{equation}
P_{\rm typ}(j) = \frac{1}{2\pi \ii}   \int_{{c}_b} \lim_{t\to \infty}  \langle e^{\lambda t^{-1/2z} J(t)} \rangle  e^{-\lambda j},
\label{inverse_Laplace}
\end{equation}
where  ${c}_b$ is a Bromwhich contour that passes to the right of all singularities of the integrand.\\

\noindent To compute the limiting FCS in \eqref{inverse_Laplace} we observe that typical charge fluctuations  are carried by typical vacancy fluctuations, $n_\pm - t \overline \rho = \mathcal{O}(t^{1/2})$, wherefore we 
introduce
\begin{equation}
n_\pm = t \overline \rho + \delta_\pm t^{1/2}.  \label{typ_n_fluc}
\end{equation}
Recalling the De Moivre-Laplace theorem we have that
\begin{align}
\rho^{2t} \sum_{n_\pm =0}^t \binom{t}{n_+} \binom{t}{n_-}\nu^{n_- + n_+}  \mu_-^{n_+ - n_-} \simeq  \iint_{\mathbb{R}^2}  \frac{\dd \delta_- \dd \delta_+}{2\pi \rho \overline \rho} e^{-\frac{\delta^2_- + \delta^2_+}{2\rho\overline \rho}} \mu_-^{(\delta_+-\delta_-) t^{1/2}},
\end{align}
and the expression \eqref{dressed_6_vertex_partition_step_eq} becomes
\begin{equation}
\langle e^{\lambda J(t)} \rangle \simeq \iint_{\mathbb{R}^2}  \frac{\dd \delta_- \dd \delta_+}{2\pi \rho\overline \rho} e^{-\frac{\delta^2_- + \delta^2_+}{2\rho\overline \rho}} \mu_-^{(\delta_+-\delta_-) t^{1/2}}  \mu_-^{n_+ - n_-}   \mathbb{E}_{\rm step}(\mu^{N_{t\rho-\delta_+t^{1/2}}}|t\rho-\delta_-t^{1/2}).
\end{equation}
Introducing the sum and difference variables 
\begin{equation}
\delta = \delta_+ - \delta_-, \qquad \kappa =  \delta_+ + \delta_-,
\end{equation}
we observe that the expectation value of the six-vertex model is asymtotically independent of $\kappa$ as follows from Eq.~\eqref{CGF_series_result}.
Integrating over $\kappa$ we have
\begin{equation}
\langle e^{\lambda J(t)} \rangle \simeq \int_{-\infty}^{\infty} \frac{\dd \delta}{\sqrt{4\pi \rho\overline \rho}} e^{-\frac{\delta^2}{4\rho\overline \rho}} \mu_-^{\delta t^{1/2}} \mathbb{E}_{\rm step}(\mu^{N_{t\rho-\delta_+ t^{1/2}}}|t\rho - \delta_- t^{1/2}).
\end{equation}
In view of the dynamical rescaling of the counting field \eqref{dynamical_rescaling} we can consider cases where it is small. We can then expand
\begin{equation}
\mu_\pm = 1 \mp b \lambda + \lambda^2/2 + \mathcal{O}(\lambda^3), \label{mu_pm_series}
\end{equation}
whence $\mu = \mu_-\mu_+ = 1 + (1-b^2) \lambda^2 + \mathcal{O}(\lambda^3)$ and we recall that $b$ is the charge per particle in the initial state. We similarly expand \eqref{CGF_series_result} for $\delta \geq 0$ to lowest order in $\lambda$
\begin{equation}
\log \mathbb{E}_{\rm step}(\mu^{N_{x}}|t) = \lambda^2(1-b^2) \sqrt{\tfrac{t}{\gamma \pi}}\ p\left(\delta\sqrt{\gamma/4}\right) + \mathcal{O}(\lambda^3).
\end{equation}
Using the reflection relation \eqref{step_reflection} to extend the result to $\delta < 0$ we find
\begin{equation}
\langle e^{\lambda J(t)} \rangle \simeq \int_{-\infty}^\infty \frac{\dd \delta}{\sqrt{4\pi \rho\overline \rho}} e^{-\frac{\delta^2}{4\rho\overline \rho}} \mu_{-{\rm sgn}\, \delta}^{|\delta| t^{1/2}}  \exp\left[ \lambda^2(1-b^2) \sqrt{\tfrac{t\rho}{\pi \gamma}}\ p \left(\sqrt{\tfrac{\gamma}{4\rho}}|\delta|\right)\right]. \label{asymp_dressed_FCS}
\end{equation}
It remains to consider typical charge fluctuations in initial measures with different average charge densities.

\subsubsection{Non-vanishing charge density}
We first consider the case of $b\neq 0$ where the lowest order in $\lambda$ comes from  the expansion \eqref{mu_pm_series} of $\mu_\pm$, indicating a rescaling  of the counting field with a ballistic dynamical exponent $z=1$ which eliminates the last term as $t \to \infty$
\begin{equation}
\lim_{t\to \infty} \langle e^{\lambda t^{-1/2} J(t)} \rangle = \int_{-\infty}^\infty \frac{\dd \delta}{\sqrt{4\pi \rho\overline \rho}} e^{-\frac{\delta^2}{4\rho\overline \rho} + b \delta \lambda} = e^{\lambda^2 b^2 \rho\overline \rho}, \qquad b \neq 0,
\end{equation}
and corresponds to a Gaussian distribution of typical fluctuations for $b\neq 0$
\begin{equation}
P_{\rm typ}(j) = \frac{1}{\sqrt{2\pi \sigma^2}} e^{-\frac{j^2}{2 \sigma^2}}, \label{gauss_typical_b}
\end{equation}
where $\sigma^2 = 2b^2\rho \overline \rho$. The distribution \eqref{gauss_typical_b} matches the single-file result of Ref.~\cite{Krajnik2022}.

\begin{figure}[h!]
\centering
\includegraphics[width=0.9\linewidth]{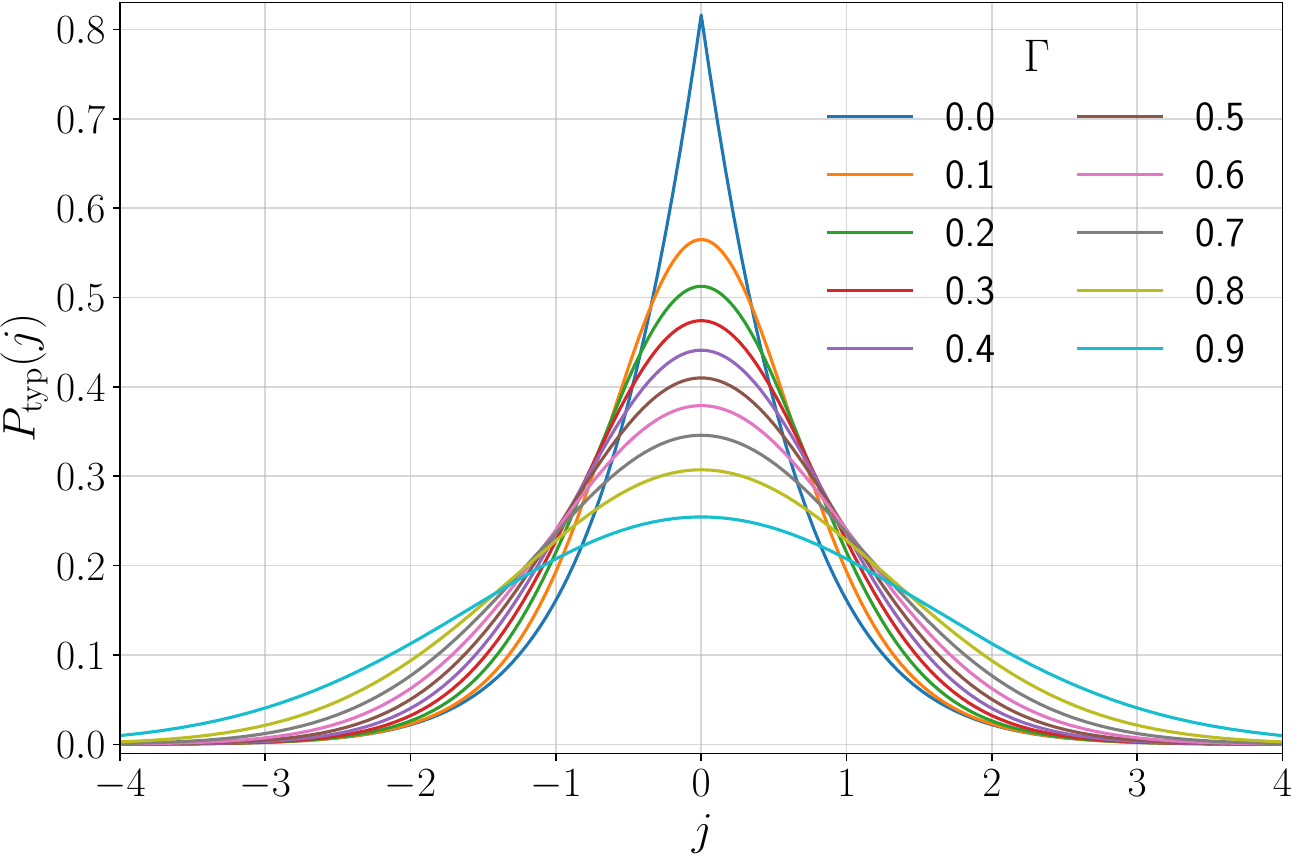}
\caption{The distribution \eqref{typical_final_hf} (plotted for $\rho = 1/2$) of typical charge fluctuations in the stochastic cellular automaton at zero net charge interpolates between a Mainardi-Wright distribution in the single-file limit ($\Gamma =0$, blue curve) and a Gaussian distribution in the free limit ($\Gamma \to 1$, teal curve).}
\label{fig:distribution_family}
\end{figure}

\subsubsection{Vanishing charge density} At zero net charge (i.e.~for $b=0$) the lowest order in $\lambda$ in the exponent of Eq.~\eqref{asymp_dressed_FCS} is $\lambda^2$, indicating a rescaling of the counting field with a diffusive dynamical exponent $z=2$ which gives for $t \to \infty$
\begin{equation}
\lim_{t\to \infty} \langle e^{\lambda t^{-1/4} J(t)} \rangle = \int_{-\infty}^\infty \frac{\dd \delta}{\sqrt{4\pi \rho\overline \rho}}e^{-\frac{\delta^2}{4\rho\overline \rho} + \frac{\lambda^2}{2}\left[|\delta| + \sqrt{\frac{4\rho}{\pi \gamma}}\, p \left(\sqrt{\tfrac{\gamma}{4\rho}}|\delta|\right) \right]},
\end{equation}
and corresponds to a non-Gaussian distribution of typical fluctuations for $b=0$
\begin{equation}
P_{\rm typ}(j) = \frac{1}{\pi \sigma }\int_{0}^{\infty} \frac{\dd u}{\sqrt{u[1+s(u/a)]}} e^{-\frac{u^2}{2\sigma^2}-\frac{j^2}{2u[1+s(u/a)]}}, \label{typical_final_hf}
\end{equation}
 where $\sigma^2 = 2  \rho \overline \rho $ while $a = 2 \sqrt{\rho/\gamma}\geq 0$ is a scale parameter
 and we have introduced the auxiliary function
 \begin{equation}
 s(z) =  \pi^{-1/2} z^{-1}e^{-z^2} - {\rm erfc}\, z.
 \end{equation}
The one-parameter family of distributions \eqref{typical_final_hf} is plotted in Figure \ref{fig:distribution_family}.
In the single-file limit $\Gamma \to 0$, we have $a\to 0$ with $s(z \to \infty)\to 0$ and \eqref{typical_final_hf} recovers the Mainardi-Wright distribution of order 1/4 \cite{Mainardi1996,Mainardi2020} derived microscopically in Ref.~\cite{Krajnik2022} and hydrodynamically in Ref.~\cite{Yoshimura2024}. In the free limit $\Gamma \to 1$ we have $a \to \infty$ with $s(z \to 0) \to \pi^{-1/2} z^{-1}$ and the distribution \eqref{typical_final_hf} approaches a Gaussian distribution with variance $2 \pi^{-1/2} \rho /\gamma$. Divergence of the variance at $\Gamma \to 1$ (i.e.~$\gamma \to 0$) indicates a change of dynamical exponent from $z=2$ to $z=1$ for free dynamics. As shown in the representative example of Fig.~\ref{fig:numerical_comparison} (see Sec.~\ref{sec:summary}), a comparison between the analytical prediction \eqref{typical_final_hf} and direct numerical simulations shows excellent agreement for all considered values of $\Gamma$.

\subsection{Large fluctuations}
\label{sec:dressed_large_asymptotics}
Large fluctuations of the charge current are characterized by the large deviation form of the full counting statistics
\begin{equation}
F(\lambda) = \lim_{t \to \infty} t^{-1}\log \langle e^{\lambda J(t)} \rangle,
\end{equation}
which is dominated by large $\mathcal{O}(t)$ vacancy fluctuations --- this follows directly from an asymptotic analysis of the `vacancy terms' in Eq.~\eqref{dressed_6_vertex_partition_step_eq}. On the other hand, as shown in Section~\ref{sec:asymp_FCS}, the asymptotic full counting statistics for $t-x = \mathcal{O}(t) \geq 0$  is exponentially close to unity and therefore asymptotically negligible compared to the vacancy terms. However, for $t-x = \mathcal{O}(t) < 0$ there is an $\mathcal{O}(t)$ contribution due to the reflection relation \eqref{step_reflection}.
We therefore have
\begin{equation}
F(\lambda) = \lim_{t\to \infty}  t^{-1}\log \left[ \rho^{2t} \sum_{n_+,n_- =0}^t \binom{t}{n_+} \binom{t}{n_-}\nu^{n_- + n_+}  \mu^{|n_+-n_-|}_{{\rm sgn}(n_--n_+)} \right], \label{equilibrium_large2}
\end{equation}
which precisely matches the single-file expression analyzed in Ref.~\cite{Krajnik2022}.  Given that Eq.~\eqref{equilibrium_large2} is independent of the crossing probability $\Gamma$ we can directly quote the result of Ref.~\cite{Krajnik2022}
\begin{equation}
F(\lambda) = \frac{1}{2}\log\left[1 + \rho \overline \rho(\mu_b + \mu^{-1}_b -2) \right], \label{LD_result}
\end{equation}
where $\mu_b = \cosh \lambda + |b| \sinh |\lambda|$. This is a further microscopic confirmation of the recent prediction of ballistic macroscopic fluctuation theory~\cite{Yoshimura2024}.

\section{Conclusions}
\label{sec:conclusion}
We have studied equilibrium charge current fluctuations in a cellular automaton with stochastic particle scattering that naturally generalizes the class of charged single-file dynamics \cite{Krajnik2022,Krajnik2024}. We mapped the full counting statistics of the cellular automaton to a vacancy-dressed full counting statistics of the bistochstic six-vertex, which we expressed as a Fredholm determinant using techniques of integrable combinatorics. The study of its asymptotics was facilitated by appropriately manipulating the multiple-integral representation of the traces of the corresponding integral kernel and a reflection relation. As an intermediate result we have obtained the asymptotic form of the full counting statistics of the bistochastic six-vertex model on a square lattice with diffusive shape fluctuations. The charge fluctuations of the cellular automaton followed from an additional analysis of the vacancy combinatorics. 

We found that at vanishing charge bias an arbitrarily small rate of back-scattering changes the dynamical exponent from ballistic to diffusive. The corresponding typical distribution is non-Gaussian and smoothly interpolates between a Mainardi--Wright distribution in the single-file limit and a Gaussian distribution in the free limit. The probability of large charge fluctuations is independent of the crossing probability, confirming a recent prediction of ballistic macroscopic fluctuation theory~\cite{Yoshimura2024}.

Our work raises several points deserving of further attention. While we have presently restricted our asymptotic analysis to equilibrium fluctuations it would be interesting to study fluctuations in generic bipartite initial states as recently done in Ref.~\cite{Krajnik2024} for the single-file limit. Likewise, using techniques similar to those employed in the present work, it is possible to study a cellular automaton with biased scattering between positive and negative particles. The hydrodynamic description of the single-file limit is given by a Euler equation with a stochastic velocity~\cite{Yoshimura2024}. It would be interesting to extend the hydrodynamic description to the present stochastic case. Lastly, fluctuations of the spin current in the easy-axis and isotropic regimes of the quantum XXZ spin-1/2 chain and its semi-classical analogue have recently been found to be anomalous \cite{Krajnik2024a,Rosenberg2024}. This calls for an extension of the integrability methods used here to that setup. More generally, such an extension could be used to prove space-time duality based conjectures~\cite{bertini2023nonequilibrium, bertini2024dynamics} on the dynamics of charge fluctuations in integrable models.  

\section*{Acknowledgements} 
We thank Enej Ilievski, Johannes Schmidt, Vincent Pasquier and Takato Yoshimura for collaboration on related topics. We thank Takato Yoshimura, Vincent Pasquier, Amol Aggarwal, Enej Ilievski for comments on the manuscript and Tomohiro Sasamoto for bringing Ref.~\cite{Imamura_Mallick_Sasamoto_2021} to our attention. 
\v{Z}K is supported by the Simons Foundation as a Junior Fellow of the Simons Society of Fellows (1141511). BB is supported by the Royal Society through the University Research Fellowship No.\ 201101. KK is supported by the Leverhulme Trust through the Early Career Fellowship No. ECF-2022-324. TP acknowledges support by European Research Council (ERC) through Advanced grant QUEST (Grant Agreement No. 101096208), and Slovenian Research and Innovation agency (ARIS) through the Program P1-0402 and Grants N1-0219, N1-0368.
\section*{References}

\bibliographystyle{iopart-num}
\bibliography{master_bibfile.bib}

\end{document}